\documentclass[%
preprint,
 amsmath,amssymb,
 aps,
pre, 
]{revtex4-1}
\usepackage[utf8]{inputenc}
\usepackage[T1]{fontenc}
\usepackage{graphicx}
\usepackage{dcolumn}
\usepackage{bm}
\usepackage[dvipsnames]{xcolor}
\definecolor{gr}{rgb}{0,0.8,0}
\usepackage{hyperref}
\usepackage{csquotes}
\usepackage{float}
\usepackage{multirow}
\usepackage{epstopdf}
\usepackage{siunitx}

\newcommand{\orcid}[1]{\href{https://orcid.org/#1}{\textcolor[HTML]{A6CE39}{\aiOrcid}}}
\DeclareMathOperator{\Tr}{Tr}

\newcommand{\Q}{\mathbf{Q}}
\newcommand{\QQ}{\mathbf{Q}}

\newcommand{\PP}{\mathbf{P}}
\newcommand{\rr}{\mathbf{r}}

\newcommand{\nn}{\mathbf{\hat n}}

\begin{document}

\preprint{APS/123-QED}

\title{Twist-bend nematic phase from Landau-de Gennes perspective}
\author{Lech Longa\href{https://orcid.org/0000-0002-4918-6518}{\includegraphics[scale=.6]{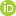}}\, 
}
\email[E--mail address:]{ lech.longa@uj.edu.pl}
\author{Wojciech Tomczyk\href{https://orcid.org/0000-0002-7649-3329}{\includegraphics[scale=.6]{orcid_16x16.png}}\,}%
\email[E--mail address:]{ wojciech.tomczyk@doctoral.uj.edu.pl}
\affiliation{Jagiellonian University, Institute of Theoretical Physics, Department of Statistical Physics, 
\L{}ojasiewicza 11, 30--348 Krak\'{o}w, Poland}

\date{\today}

\begin{abstract}
The understanding of self-organization in 
the twist-bend nematic ($N_\text{TB}$) phase, identified in 2011 
in liquid crystal dimers, is at the forefront of soft matter 
research worldwide.  
This new nematic phase develops structural chirality in the isotropic ($I$) 
and the uniaxial nematic ($N_\text{U}$) phases, despite the fact 
that the 
molecules forming the structure are chemically \emph{achiral}. Molecular, shape-induced flexopolarization provides a viable mechanism
for a qualitative understanding of $N_\text{TB}$ and the 
related phase transitions. The key question that remains is whether 
with this  mechanism one can also explain quantitatively the presently 
existing experimental data. To~address this issue we    
propose a generalization of the mesoscopic 
Landau-de Gennes theory of nematics, where higher-order elastic terms 
of the alignment tensor are taken into account, in addition to the 
lowest-order flexopolarization coupling.  
The theory is not only capable of explaining  the 
appearance of $N_\text{TB}$ but also stays in quantitative agreement with 
experimental data. In exemplary calculations, we take the data
known for CB7CB flexible dimer - the ``drosophila fly'' in the studies 
of $N_\text{TB}$ [A.~J\'{a}kli et~al., Rev. Mod. Phys. 90, 045004 (2018)]
- and estimate the constitutive 
parameters of the model from temperature variation of the nematic 
order parameter and the Frank elastic constants in the nematic phase. 
Then we  seek for  relative stability and properties 
of the isotropic,  uniaxial nematic and twist-bend nematic phases.
In particular, we evaluate various properties of $N_\text{TB}$,
like temperature variation of the structure's wave vector, conical angle,
flexopolarization, and remaining order parameters. 
We also look into the fine structure of $N_\text{TB}$,
like its biaxiality - the~property, which is difficult to access 
experimentally at the nanoscale. 
\end{abstract}

\pacs{Valid PACS appear here}% PACS, the Physics and Astronomy
\maketitle
\section{Introduction }
%%%%%%%%%%%%%%%%%%%%%%%%%%%
%
%
Undoubtedly, the short-pitch heliconical structure formed by an 
ensemble of achiral bent-core-like mesogens and commonly referred to as the
nematic twist-bend,  is one of the most astonishing liquid crystalline 
phases. It is the first example in nature of a structure 
where mirror symmetry is spontaneously broken without any support 
from a long-range positional order. The structure itself is 
a part of an over 130-year-old tradition of liquid crystal science demonstrating 
that even a minor change in the molecular chemistry can 
lead to a new type of liquid crystalline order, which differs 
in the degree  of orientational and 
translational self-organization, ranging from molecular 
to macro scales \cite{deGennesBook,RevModPhys.82.897,RevModPhys.90.045004}.

The most common of all known liquid crystalline phases is 
the uniaxial nematic phase ($N_\text{U}$), where anisotropic molecules 
or molecular aggregates orient, on the average, parallel to each other.
Their local, mean orientation at the point $\tilde{\rr}$ of coordinates
($\tilde{x},\tilde{y},\tilde{z}$) is described by a single 
mesoscopic direction $\nn(\tilde{\rr})$ ($|\nn(\tilde{\rr})|=1$), 
known as the director.
Due to the statistical head-to-tail inversion symmetry of the local molecular
arrangement the director states  $\nn(\tilde{\rr})$ and
$-\nn(\tilde{\rr})$  are equivalent.
With an inversion symmetry and with a rotational symmetry of molecular 
orientational distribution 
about $\nn(\tilde{\rr})$ the existence of the director 
is a basic property that distinguishes 
the uniaxial nematics from an ordinary isotropic liquid.
That is,  the $N_\text{U}$ phase is a non-polar 3D liquid with 
long-range orientational order characterized by 
${\cal{D}}_{\infty h}$ point group symmetry. 

One important consequence of $\nn(\tilde{\rr})$ being indistinguishable from
$-\nn(\tilde{\rr})$ is that the 
primary order parameter of the uniaxial nematics is the second-rank ($3\times3$) traceless
and symmetric alignment tensor (the  quadrupole moment of the local angular
distribution function of the molecules' long axes)
\begin{equation}\label{alignmentTensor}
   \tilde{ \QQ}_U(\tilde{\rr}) = \tilde{ S}\left(\nn(\tilde{\rr}) 
   \otimes \nn(\tilde{\rr}) -\frac{1}{3} \mathbf{I}\right)
\end{equation}
having components $ \tilde{Q}_{U,\alpha\beta}$;
$\tilde{S}$ is the scalar order parameter describing the degree of (local)
molecular orientational ordering along $\nn(\tilde{\rr})$ and
 $\mathbf{I}$ denotes the identity matrix.

Beyond conventional uniaxial nematics further nematic liquid phases, that (by
definition) have only short-ranged positional ordering,  were recognized. They
involve $\mathcal{D}_{2h}$ symmetric  biaxial nematics ($N_\text{B}$)  for
non-chiral materials and  cholesteric $(N^\ast)$ along with blue phases (BP) for chemically chiral
mesogens, characterized locally by $\mathcal{D}_{2}$ point group symmetry.
In order to account for their local orientational order 
we need a full, symmetric and traceless alignment tensor
$\tilde{\QQ}(\tilde{\mathbf{r}})$ with three different eigenvalues,
as opposed to the uniaxial nematic ($\ref{alignmentTensor}$),
where only two eigenvalues of $\QQ \equiv \QQ_U$ are different. 
 
This four-members nematic family is ubiquitous in nature and 
it has not been expanding for many years
\cite{deGennesBook}. However, very recently the situation has changed
with important discovery  of two fundamentally 
new nematics:  the twist-bend nematic phase $(N_\text{TB})$
\cite{Panov2010,Cestari2011,Borshch2013,Chen2013}
and the nematic splay phase $(N_S)$ \cite{PhysRevX.8.041025}, and it seems these
discoveries only mark a beginning of new, fascinating  research direction 
in soft matter science 
\cite{RevModPhys.90.045004,Tuchband10698,UK2018,Pajak2018,LongaPajak2016}. 

Without any doubt the discovered $N_\text{TB}$ phase is different
than  3D liquids known to date, because it exhibits a macroscopic chirality, 
while formed from chemically achiral, bent-core-like molecules.
A direct manifestation of chirality is an  average orientational 
molecular order that forms a local helix with a pitch spanning from several to over a dozen of nanometers, 
in the absence of any long-range positional order of molecular centers of mass.
$N_\text{TB}$ is stabilized as a result of (weakly) first order
phase transition from the uniaxial nematic phase, or  directly from
the isotropic phase \cite{ArchboldDavis2015,DawoodGrossel2016}
and therefore (as already mentioned) its emergence is probably 
one of the most unusual manifestation of mirror symmetry breaking 
(SMSB) in three-dimensional liquids.

\begin{figure}
    \centering
    \includegraphics[width=.95\textwidth]{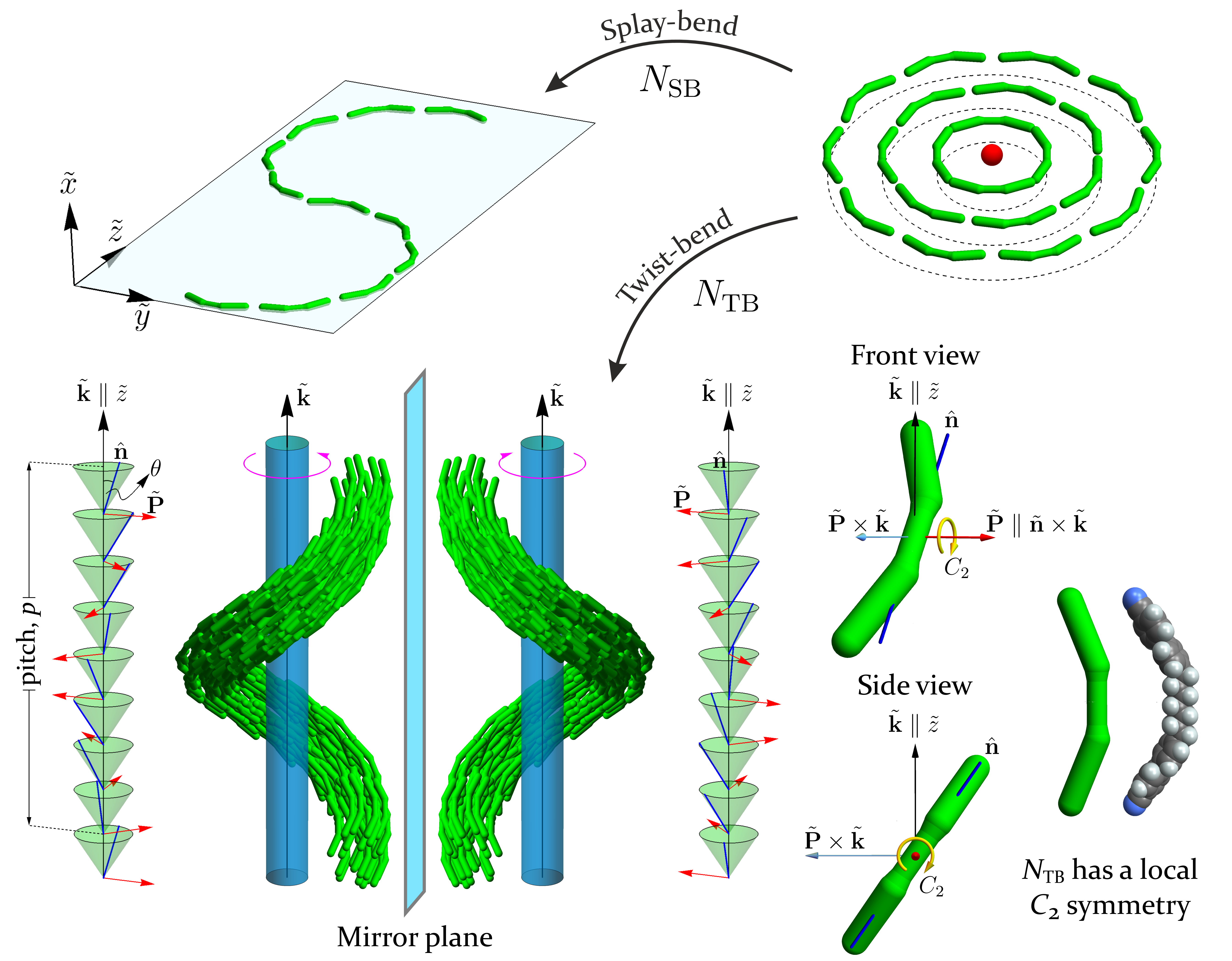}
    \caption{ Pure bend distortion in 2D leads to the emergence of defects 
    (red sphere). Their appearance can be circumvented by alternating 
    the bend direction periodically 
     or allowing nonzero twist by lifting bend  into the third dimension. 
     These possibilities, respectively, 
    give rise to the two alternative  nematic  ground  states: 
    splay-bend ($N_\text{SB}$) and twist-bend ($N_\text{TB}$). 
    Twist-bend nematic has been firstly observed in the phase 
    sequence of liquid crystal dimer, 
    1$^{\prime\prime}$,7$^{\prime\prime}$-bis(4-cyanobiphenyl-4'-yl)heptane 
    (CB7CB), where two identical cyanobiphenyl mesogenic groups are linked 
    by a heptane spacer (CB7CB molecule can be viewed as 
    having three parts: two identical rigid end groups connected 
    by a flexible spacer). Schematic representation of molecular 
    organization in the $N_\text{TB}$ with right and left handedness 
    (ambidextrous chirality) has been depicted at the bottom of the image. 
    Right/left circular cone of conical angle $\theta$ shows the tilt 
    between the director $\mathbf{\hat{n}}$ and the helical symmetry 
    axis, parallel to the wave vector $\tilde{ \textbf{k}}$. Red arrow represents 
    polarization $\tilde{ \textbf{P}}$, where $\tilde{ \textbf{P}}$\,$\parallel$\, 
   $\tilde{ \textbf{k}}$\,$\times$\,$\mathbf{\hat{n}}$; black arrow 
    is the direction of $\tilde{\textbf{k}}$. Note that $N_\text{TB}$ has a local 
    $\mathcal{C}_{2}$ symmetry with a two-fold symmetry axis around $\tilde{ \textbf{P}}$.} 
    \label{fig:negative_bend}
\end{figure}

At the theoretical level a possibility of SMSB in bent-shaped mesogens has been suggested 
by Meyer already in 1973. He pointed out  that bend deformations, which should 
be favored by bent--shaped molecules, might lead to flexopolarization--induced 
chiral structures \cite{MeyerFlexopolarization}.
 About thirty years later Dozov \cite{Dozov2001} considered the Oseen-Frank (OF)  free energy 
$F_{OF}= \tilde{V}^{-1}{\int_{\tilde{V}}}\, f_{\text OF}\, \mathrm{d}^{3}{ { \tilde{\rr}} }$ 
of the director field ${\nn}(\tilde{\rr})$ \cite{TF9332900883,DF9582500019}, where
\begin{equation}\label{Oseen-Frank}
    f_{\text OF}=\frac{1}{2}[K_{11}(\tilde{\nabla}\cdot {\nn})^2+
    K_{22}({\nn}\cdot\tilde{\nabla}\times{\nn})^2
    +K_{33}({\nn}\times\tilde{\nabla}\times{\nn})^2],
\end{equation}
and where $K_{11}$, $K_{22}$ and $K_{33}$ are splay, twist, and bend elastic constants, 
respectively.  He correlated  a possibility of 
SMSB in nematics with  the sign change of the  bend  elastic constant, $K_{33}$.
 In this latter case, in order to guarantee the existence of a stable ground state, 
some {higher order elastic terms} had to be added to $f_{\text OF}$.
Limiting to defect-free structures, Dozov predicted competition between 
a twist--bend nematic  phase,
where the director  simultaneously twists and bends 
in space by precessing on the side of a right circular cone
and a planar splay-bend phase with alternating domains of splay and bend, both shown in  
Fig.~\ref{fig:negative_bend}. If we take into account temperature dependence 
of the Frank elastic constants then the uniaxial nematic phase becomes 
unstable to the formation of modulated structures at $K_{33}=0$, which 
is the critical point of the model. The behaviour of the system depends on
the relationship between the splay and twist elastic constants.
As it  turns out the twist-bend ordering wins if 
$K_{11} > 2 K_{22}$, while the splay-bend phase is more stable if $K_{11} < 2 K_{22}$.
Assuming that  the wave vector $\tilde{\mathbf{k}}$ of $N_\text{TB}$ 
stays parallel to
the $\mathbf{\hat{z}}$--axis of the laboratory system of frame
($\tilde{\mathbf{k}}=\tilde{k} \mathbf{\hat{z}}$)  the 
symmetry-dictated,  gross features of the heliconical 
$N_\text{TB}$ structure are essentially accounted for by the uniform
director modulation
\begin{equation}\label{nn}
 \mathbf{\hat{n}}(\tilde{z}) = \mathcal{R}_{\mathbf{\hat{z}}}(\phi) \mathbf{\hat{n}}(0)=
 [\cos (\phi) \sin (\theta), \sin (\phi) \sin (\theta), \cos (\theta)],
\end{equation}
where  $\mathbf{\hat{n}}(0)=[\sin (\theta),0, \cos (\theta)]$ and where
$\mathcal{R}_{\hat{z}}(\phi) = \mathcal{R}_{\hat{z}}(\phi(\tilde{z}))$ 
is the homogeneous rotation about $\mathbf{\hat{z}}$
through the azimuthal angle  $\phi= \pm \tilde{k} \tilde{z} = \pm 2\pi \tilde{z}/p$ and where 
$p$  is the pitch. The $\pm$ sign
indicates that both left--handed and right--handed chiral domains should
form with the same probability, which is manifestation  of SMSB in the bulk.
Note that the molecules in $N_\text{TB}$ are inclined, on the average, 
from  ${\tilde{\mathbf{k}}}$ by the conical angle $\theta$ --
the angle between $\mathbf{\hat{n}}$ and
the wave vector $\tilde{\mathbf{k}}$ (Fig. \ref{fig:negative_bend}). 
The symmetry of $N_\text{TB}$ also implies that   
the structure must be locally polar with the
polarization vector, $\tilde{\mathbf{{P}}}$, 
staying perpendicular both to the director and the wave vector
\begin{equation}\label{pp}
 \tilde{\mathbf{{P}}}(\tilde{z}) = \mathcal{R}_{\mathbf{\hat{z}}}(\phi) 
 \tilde{\mathbf{P}}(0)= \tilde{p}_0\,[\sin (\phi),-\cos (\phi), 0].
\end{equation}
Hence, in the nematic twist-bend phase, both $\hat{\mathbf{n}}$ 
and $\tilde{\mathbf{P}}$ rotate along the
helix direction ${\tilde{\mathbf{k}}}$ giving rise to a phase with 
constant bend and twist deformation of no mass density modulation (Fig. \ref{fig:negative_bend}).

In 2013 Shamid \emph{et~al.} \cite{Shamid&Dhakal&Selinger2013} 
developed a Landau theory for bend flexoelectricity and showed
that the results of Dozov are in line with Meyer's idea of flexopolarization-induced
$N_\text{TB}$. Their theory predicts a continuous $N-N_\text{TB}$ transition, where
the effective bend elastic constant, renormalized by the flexopolarization coupling,  
changes sign for sufficiently large coupling. The
corresponding structure develops modulated polar order, 
averaging to zero globally as in Eq.~(\ref{pp}).
The Dozov's model is also supported by measurements of 
anomalously small bend elastic constant (compared to the splay and twist  
elastic constants) in the nematic phase of materials exhibiting $N_\text{TB}$ 
(\emph{see e.g.} measurements for CB7CB dimer of Babakhanova {\emph{et al.}}
\cite{Babakhanova2017,Cukrov2017,PhysRevE.85.012701}).

The second most widely used continuum model to characterize orientational properties 
of nematics is the  minimal coupling, $SO(3)$-symmetric Landau-de Gennes (LdeG) expansion  
in terms of the local alignment tensor. It allows not only to account for a fine 
structure of inhomogeneous nematic phases, but also shows important 
generalizations of the director’s description in dealing with orientational
degrees of freedom (\emph{see e.g.} \cite{Garland2018}).
In a series of papers \cite{Longa&Trebin1990,LongaPajak2016,Pajak2018}, 
co-authored by one of us, we developed an extension of the LdeG theory
to include flexopolarization couplings. The extended theory predicted that 
the flexopolarization mechanism can make
the  $N_\text{TB}$  phase absolutely stable within the whole family of one-dimensional 
modulated structures \cite{Pajak2018}.  A qualitatively 
correct account of experimental observations in $N_\text{TB}$ (\emph{see e.g.} 
\cite{RevModPhys.90.045004}) was obtained, like trends in temperature variation of the helical
pitch and conical angle, and behaviour in the external electric field \cite{Kocot}.
The theory also predicted weakly first order phase 
transitions from the isotropic and nematic to the nematic twist-bend phase,
again in agreement with experiments \cite{Mandle2015,DawoodGrossel2016}.
Despite this qualitative success of the LdeG modelling one important
theoretical issue still left unsolved is that associated with the elastic behaviour
of the uniaxial nematic phase for  materials with stable $N_\text{TB}$. 
A few existing measurements of all three elastic constants 
in the  $N_\text{U}$ phase show that $K_{11} \gtrsim 2 K_{22}$ ($K_{22}\approx3-4\,\text{pN}$), 
while $K_{33}\approx 0.4\,\text{pN}$  near the transition into $N_\text{TB}$ \cite{Babakhanova2017}. 
That is, the splay elastic constant is about 20 times larger 
than the bend elastic constant. On the theoretical side, the LdeG expansion with only  
two distinct bulk elastic terms cannot explain this anomalously large disparity in the values 
of $K_{11}$ and $K_{33}$. Actually, it predicts that they both are 
equal in the Oseen-Frank limit \cite{BerremanMeiboom1984,LongaMonselesanTrebin1987}, 
where the alignment tensor is given by Eq.~(\ref{alignmentTensor}).
Therefore, there are anomalously small bend and splay Frank elastic constants  
 on approaching $N_\text{TB}$ in the LdeG model with flexopolarization 
\cite{LongaPajak2016}. Though this prediction might suggest 
a dominance of the structures with splay-bend deformations over the 
twist-bend ones, the $N_\text{TB}$ phase, as already discussed before, can still 
be found to be more stable than any of one-dimensional periodic structures,
including the nematic splay-bend phase \cite{Pajak2018}.
Most probably this is due to the remarkable (and unique) feature of $N_\text{TB}$ 
of being uniform everywhere in space that makes the $SO(3)$-symmetric elastic free energy
density independent of space variables \cite{Pajak2018}.
  
Central to quantitative understanding of $N_\text{TB}$ and related phase transitions is then  a construction of 
a generalized LdeG theory that releases the $K_{11}=K_{33}$ restriction 
of the minimal coupling model and accounts
for the experimental behaviour of the Frank elastic constants 
in the vicinity of $N_\text{U}-N_\text{TB}$ phase transition.
We expect that such a theory will  allow for a  systematic 
study of mesoscopic mechanisms that can be responsible 
for chiral symmetry breaking in nematics. It will also give 
a new insight into conditions that can potentially lead 
experimentalists to a discovery of new 
nematic phases. Although the choice of strategy  
has already been worked out in the literature 
\cite{Longa1986,PhysRevA.39.2160,Longa&Trebin1990} 
the main problem lies in a huge number of elastic invariants 
in the alignment tensor, contributing to the generalized elastic free
energy density of nematics. Here we show how the problem
can be solved in a systematic way if we start from 
a theory which holds without limitations 
for arbitrary one-dimensional periodic 
distortions of the alignment tensor. 
These distortions  form the most interesting class 
of structures for it obeys the recently discovered new nematic phases.
An additional requirement for the generalized LdeG theory is that 
its ground state in the absence of flexopolarization 
should be that corresponding to a constant tensor field $\tilde{\QQ}$. 
The theory so constructed will then be applied to characterize 
properties of $N_\text{TB}$ formed in the class of CB7CB-like
dimers and its constitutive 
parameters will be estimated from experimental data known for the
CB7CB dimers in the $N_\text{U}$ phase.  

This paper is organized as follows.  
In Section II  a tensor representation 
for $N_\text{TB}$ and classification of  all homogeneously deformed nematic phases
is introduced. 
In Sections III and IV  a generalized Landau-de Gennes 
theory is developed for nematic structures that are periodic only in one 
spatial direction.
In Section V  bifurcation scenarios to homogeneously deformed nematics are given. 
In  Sections VI and VII  the theory is confronted 
with experimental data for CB7CB. In particular, some estimates of 
constitutive parameters of the theory are found from the data for CB7CB 
in the $N_\text{U}$ phase. Then the theory is used to study
properties of $N_\text{TB}$ along with an extensive comparison of 
the results with available 
experimental data. Predictions are also given for the order parameters 
and the degree of biaxiality of $N_\text{TB}$. The paper is concluded with 
final remarks in Section VIII.  
\section{Alignment tensor representation for homogeneously deformed nematic phases}
In the $N_\text{TB}$ phase the director $\hat{\mathbf{n}}$  
and the polarization vector $\tilde{\mathbf{P}}$ are given by 
Eqs.~(\ref{nn}) and (\ref{pp}), while
the equivalent alignment tensor order parameter, $\tilde{\QQ}_{U,TB}$, 
is obtained by substituting (\ref{nn}) into (\ref{alignmentTensor}).
Though these models seem to account for  
gross features of orientational order observed in $N_\text{TB}$ 
they do not exhaust possible  nematic
structures that can fill space with  twist,  bend and splay.
A full spectrum of possibilities is obtained by studying an expansion
of the biaxial alignment tensor $\tilde{\mathbf{Q}}$ and the polarization field  
$\tilde{\mathbf{P}}$ in spin tensor modes of $L=2$  and $L=1$, respectively, 
and in plane waves \cite{LongaPajak2016}. Within this huge family of states
an important class of nematic structures is represented by 
uniformly  deformed states (UDS) where the  
elastic, $SO(3)$-symmetric invariants contributing to 
the elastic free energy density of nematics
\cite{PhysRevA.39.2160,Longa&Trebin1990}  
are constant in space.
For such structures the same tensor and 
polarization landscape is seen everywhere in space. 
They are periodic in,  at most, one spatial 
direction, say $\tilde{z}$,  and fill uniformly space without defects.
In analogy to the conditions (\ref{nn}) and (\ref{pp}) 
for $\nn$ and $\PP$,  they are generated  
from the tensors $\tilde{\QQ}(0)$ and $\tilde{\PP}(0)$
for $\tilde{z}=0$ by the previously defined homogeneous rotation 
$\mathcal{R}_{\hat{z}}(\pm \tilde{k} \tilde{z})\equiv \mathcal{R}_{\hat{z}}(\phi)$ 
\cite{Pajak2018,LongaLC2018}.
More specifically
\begin{eqnarray}\label{rotation}
\mathcal{R}_{\hat{z}}(\pm \tilde{k} \tilde{z}) \tilde{\QQ}(0) 
&=& \tilde{\QQ}(\tilde{z}), \nonumber\\
\mathcal{R}_{\hat{z}}(\pm \tilde{k} \tilde{z}) \tilde{\PP}(0) 
&=& \tilde{\PP}(\tilde{z}),
\end{eqnarray}
where $\pm$ labels left-- ($+$) and right--handed ($-$) heliconical structures.
Hence,  the most general representations for UDS that
generalizes Eqs.~(\ref{nn}) and (\ref{pp}),
can be cast in the form \cite{Pajak2018,LongaLC2018}:
\begin{eqnarray}
\tilde{\QQ} (\tilde{z})&=&
\frac{\tilde{x}_{0}}{\sqrt{6}} 
\left[
\begin{array}{ccc}
 -1 & 0 & 0 \\
 0 & -1 & 0 \\
 0 & 0 & 2 \\
\end{array}
\right]
+
\frac{\tilde{r}_{\pm 1}}{\sqrt{2}}
\left[
\begin{array}{ccc}
 0 & 0 & - c_{\pm 1}   \\
 0 & 0 & s_{\pm 1}  \\
 - c_{\pm 1}  & s_{\pm 1}  & 0 \\
\end{array}
\right]  
% 
%\nonumber \\ 
% &&
% \nonumber \\ 
% && \hspace{3cm} 
+
\frac{\tilde{r}_{\pm 2}}{\sqrt{2}} \left[
\begin{array}{ccc}
 c_{\pm 2}   & -s_{\pm 2}  & 0 \\
 -s_{\pm 2}  & -c_{\pm 2}   & 0 \\
 0 & 0 & 0 \\
\end{array}
\right]
,
\nonumber \\ 
 &&
\label{Q_tensor}
\end{eqnarray}
\begin{equation}
\tilde{\PP} (\tilde{z})=
 \tilde{p}_{\pm 1}
\left[
\begin{array}{c}
 -\cos \left(\pm \tilde{k} \tilde{z}+\phi _{\pm p}\right) \\ 
 \sin \left(\pm \tilde{k} \tilde{z}+\phi _{\pm p}\right) \\ 0 \\
\end{array}
\right] +
\tilde{v}_{0}
\left[
\begin{array}{c}
 0 \\ 0 \\ 1 \\
\end{array}
\right],
\label{P_tensor}
\end{equation}
where
$c_{\pm m}=\cos \left(\pm m \tilde{k}\tilde{ z}+\phi _{\pm m}\right)$ and 
$s_{\pm m}=\sin \left(\pm m \tilde{k} \tilde{z}+\phi _{\pm m}\right)$ 
and where nine real parameters
$\tilde{x}_0$, $\pm\tilde{k}$, $\tilde{r}_{\pm i} \ge 0 $, $\tilde{p}_{\pm 1}\ge 0 $, 
$ \tilde{v}_0  $, $\phi _{\pm m} $ and $\phi _{\pm p}$ 
 for each of the $\pm$ labels  characterize the fine structure of the phases, 
 especially its biaxiality.  
\begin{figure}
    \centering
    \includegraphics[width=\textwidth]{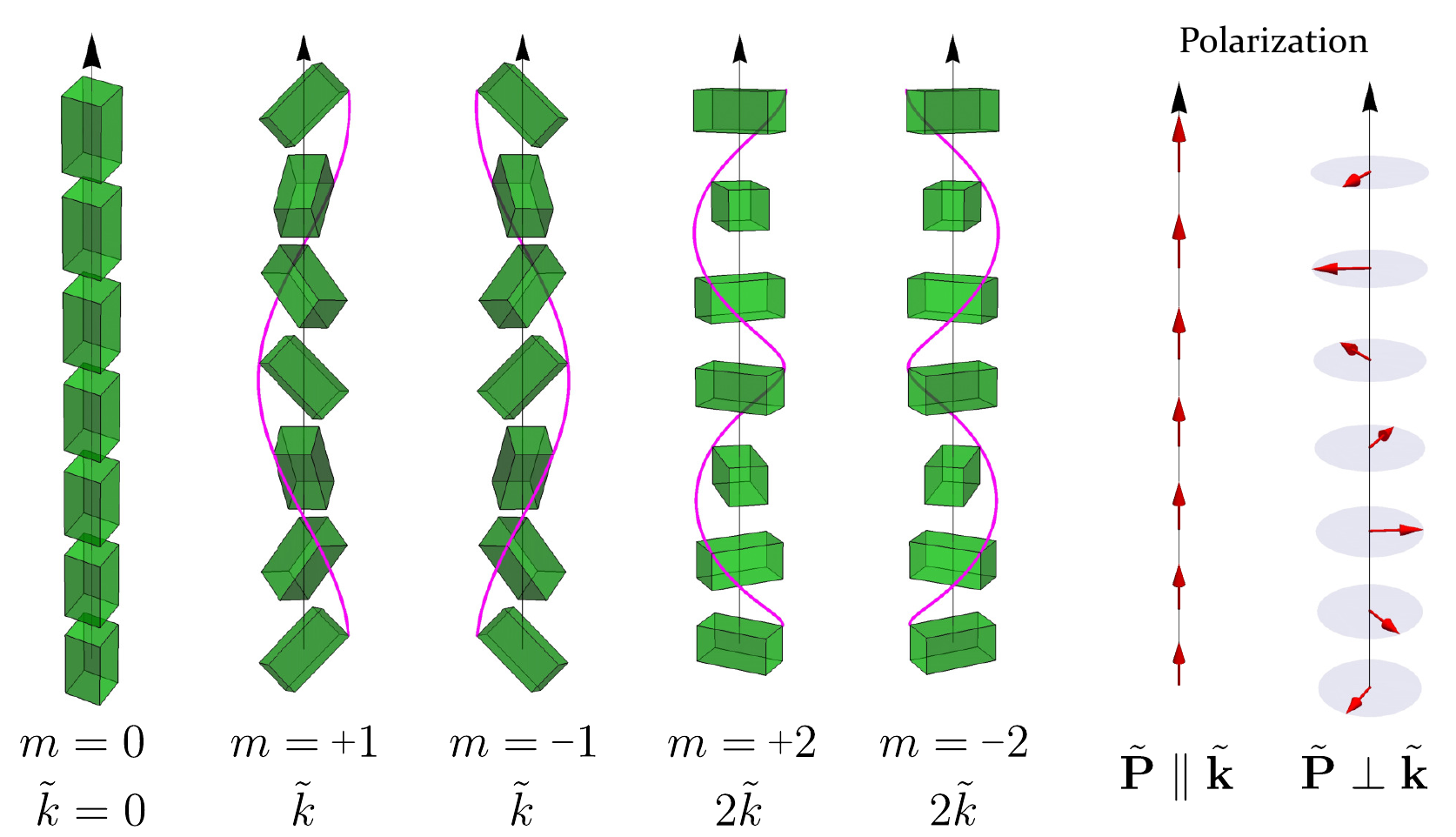}
    \caption{ Visualization of helicity modes: $m=0$, $m=\pm 1$ and $m=\pm 2$
    (change of  $m$ into $-m$ corresponds to replacement of $\tilde{\mathbf{k}}$ by
    $-\tilde{\mathbf{k}}$). Bricks represent the tensor $\tilde{\mathbf{Q}}(\tilde{\rr})$
    where the eigenvectors of $\tilde{\mathbf{Q}}(\tilde{\rr})$ are parallel to their arms, 
    while the absolute values of eigenvalues are their lengths. Red arrows represent 
    the polarization field $\tilde{\mathbf{P}}(\tilde{\rr})$.
    }
    \label{fig:my_label}
\end{figure}
Indeed, an arbitrary symmetric and traceless tensor field $\tilde{\QQ}$ 
fulfills the inequalities (\emph{see discussion in}  \cite{AllenderLonga&2008}): 
\begin{equation}\label{traceInequality}
 -1 \le w=\frac{\sqrt{6} \Tr(\tilde{\mathbf{Q}}^3)}{\Tr(\tilde{\mathbf{Q}}^2)^{\frac{3}{2}}}
  \le 1,
\end{equation}
which are satisfied as  equalities for locally
prolate $(w=1)$ and oblate $(w=-1)$ uniaxial phases. 
States of nonzero biaxiality are realized for $-1<w<1$, 
with maximal biaxiality corresponding to $w=0$. 
In particular, the  parameter
$w(\tilde{\QQ}(\tilde{z}))$ for $\tilde{\QQ}(\tilde{z})$ given by
Eq.~(\ref{Q_tensor}) reads
\begin{equation}\label{wQz}
 w(\tilde{\QQ}(\tilde{z}))= \frac{\frac{3}{2} \tilde{r}_{\pm 1}^2 
 \left(\sqrt{3} \tilde{r}_{\pm 2} \cos \left(2 \phi_{\pm 1}-\phi
   _{\pm 2}\right)+\tilde{x}_0\right)-3 \tilde{r}_{\pm 2}^2 \tilde{x}_0+ 
   \tilde{x}_0^3}{\left(\tilde{r}_{\pm 1}^2+\tilde{r}_{\pm 2}^2+
   \tilde{x}_0^2\right){}^{3/2}}.  
\end{equation}
Note that  in agreement with definition (\ref{rotation}), the parameter 
$w(\tilde{\QQ}(\tilde{z}))$ is position-independent and  
can take arbitrary value within the allowed  $[-1,1]$ 
interval, Eq.~(\ref{traceInequality}).
In contrast, for the uniaxial tensor $\tilde{\QQ}_{U,TB}$, corresponding to 
$\tilde{x}_0=\frac{\sqrt{6} S}{12} (1 + 3 \cos(2 \theta))$, 
$\tilde{r}_{\pm 1}= \frac{\sqrt{2} S}{2} \sin(2 \theta) $
and
$\tilde{r}_{\pm 2}= \frac{\sqrt{2} S}{2} \sin( \theta)^2 $
the parameter $w(\tilde{\QQ}_{U,TB})=\text{Sign}(S) = \pm 1$
($\theta$ is the conical angle).

We should mention that the 
fields in Eqs.~(\ref{Q_tensor}) and 
(\ref{P_tensor}) are insensitive to choice of
the origin of the laboratory system of frame, which allows 
to eliminate one of the phases  $\phi_{\pm i}$ ($i=1,2,p$), 
independently for each of the two states with  \enquote{$+$} 
and \enquote{$-$} subscripts. The~coefficients in 
Eqs.~(\ref{Q_tensor}) and (\ref{P_tensor}) are chosen such that the norms squared
of the order parameters are sums of squares of the coefficients:
$\Tr(\tilde{\QQ}^2)=\tilde{x}_0^2+ \tilde{r}_{\pm 1}^2+\tilde{r}_{\pm 2}^2$ and
$\Tr(\tilde{\PP}^2) = \tilde{p}_{\pm 1}^2 + \tilde{v}_0^2$. 
Together, $\tilde{\QQ}$ and $\tilde{\PP}$  characterize a family 
of all defect-free homogeneously distorted (polar) helical/heliconical 
nematic phases (they are gathered in Table~\ref{tab:structures}). 
\begingroup
\setlength{\tabcolsep}{10pt}
\renewcommand{\arraystretch}{1.5} 
\begin{table}
\caption{Family of  uniformly deformed nematic structures that can be constructed 
out of the  fields $\tilde{\QQ}$  and  $\tilde{\PP}$. Limiting  cases of 
constant $\tilde{\QQ}$ and $\tilde{\PP}$ are also included. 
\vspace{.2cm}}
{\begin{tabular}{|c|c|c|}\hline
&&\\
\textbf{Structure } & \textbf{Nonzero amplitudes} & \textbf{Abbreviation} \\
&& \\
\hline
\multicolumn{3}{|c|}{\textbf{Nonpolar structures}} \\ \hline
(a) uniaxial nematic  & $\tilde{x}_{0}$ &  $N_\text{U}$ \\
(b) biaxial nematic & $\tilde{x}_{0}$,  $\tilde{r}_1$, $\tilde{r}_2$, 
     $\tilde{k}\to 0$   & $N_\text{B}$\\
(c) (ambidextrous) cholesteric & $\tilde{x}_{0}$,  $\tilde{r}_2$, $\tilde{k}_{N^\ast}=2 
\tilde{k} \ne 0$  & $N_C$ \\ \hline
\multicolumn{3}{|c|}{\textbf{Locally polar structures}} \\ \hline
(d) locally polar cholesteric & as in (c), $\tilde{p}_1$  & $N_{Cl}$ \\
(e) nematic twist--bend  & $\tilde{x}_{0}$, $\tilde{r}_1$,  $\tilde{r}_2$, 
$\tilde{p}_1$, $\tilde{k} \ne 0$   & $N_\text{TB}$ \\ \hline
\multicolumn{3}{|c|}{\textbf{Globally polar structures}} \\ \hline
{(f) polar (a)--(e)} & {any of (a)--(e), $\tilde{v}_0$} &  add subscript 
\enquote{$p$} to (a)--(e) \\ \hline 
\end{tabular}}
\label{tab:structures}
\end{table}
\endgroup
\section{Generalized Landau-de Gennes expansion for 1D periodic nematics}
In this section we introduce a generalized Landau-de
Gennes free energy (GLdeG) expansion in  $\tilde{\mathbf{Q}}$ and  $\tilde{\mathbf{P}}$  
capable of quantitative description of the systems with stable one-dimensional periodic 
nematics. The most important members of this family are the nematic twist-bend phase 
\cite{RevModPhys.90.045004} and recently discovered nematic splay phase \cite{PhysRevX.8.041025}.  
Our main effort in this and next section will concentrate on general characterization 
of GLdeG. An example of spatially homogeneous structures with its prominent representative 
- the $N_\text{TB}$ phase - will be studied  in great detail. 
Parameters entering the LdeG expansion will be estimated  
from experimental data for the CB7CB compound in the uniaxial
nematic phase. Then, the properties of the $N_\text{TB}$ phase resulting 
from so constructed GLdeG expansion will be calculated  
and compared with available experimental data. 

We assume that the stabilization
of $N_\text{TB}$ is due to  entropic, excluded volume flexopolarization
interactions \cite{GrecoFerrarini2015}, induced by sterically 
polar molecular bent cores.
The direct interactions between electrostatic dipoles  
will be disregarded \cite{GrecoFerrarini2015} and long-range  polar 
order will be attributed to  the molecular shape polarity. 
With  $\tilde{\mathbf{Q}}$ and  $\tilde{\mathbf{P}}$
the general LdeG expansion then reads \cite{LongaMonselesanTrebin1987,Longa&Trebin1990}
\begin{equation}\label{eq:freeEnergy}  {
 \tilde{ F}=\frac{1}{\tilde{ V}}\int_{\tilde{ V}}\, \tilde{ f}_{tot}\, \mathrm{d}^{3}{ { \tilde{\rr}} }
    = \frac{1}{\tilde{V}}\int_{\tilde{ V}}\left(\tilde{ f }_{b,Q} +
    \tilde{ f }_{e,Q} + \tilde{ f}_{P}+
    \tilde{ f}_{QP} % + \tilde{ f}_{E}
    \right)
    \mathrm{d}^{3}{\tilde{\rr}},}
\end{equation}
where $\tilde{\rr}$ is the position vector, $\tilde{V}$ is the system's volume and 
where the free energy densities,  $\tilde{ f}_{x,X}$
$\tilde{ f}_{X}$,  are constructed out of the
fields $X$. They involve the bulk nematic part $\tilde{ f }_{b,Q}$, the nematic
elastic part $\tilde{ f }_{e,Q}$ and the parts $\tilde{ f}_{P}$ and
$\tilde{ f}_{QP}$ responsible for the onset of chirality in the nematic phase. 
Although the general theory has plenty of constitutive parameters 
part of them, at least for CB7CB,  can be estimated from existing 
experimental data for the $N_\text{U}$ and at the $I-N_\text{U}$ and $N_\text{U}-N_\text{TB}$ phase transitions.
One of the issues we would like to understand is whether the theory 
so constructed allows to account for the quantitative properties 
of the nematic twist-bend phase, below $N_\text{U}-N_\text{TB}$ phase transition.
%
%%%%%%%%%%%%%%%%%%%%%%%
%
%%%%%% BULK PART %%%%%%
%
%%%%%%%%%%%%%%%%%%%%%%%
%
\subsection{Bulk nematic free energy}\label{BNP}
According to the phenomenological Landau-de Gennes (LdG) theory the equilibrium 
bulk properties of nematics can be found from a nonequilibrium
free energy, constructed as an $SO(3)$-symmetric
expansion in powers of $\tilde{\mathbf{Q}}$. There are only two types of independent 
$SO(3)$ invariants that can be constructed out of $\tilde{\mathbf{Q}}$, 
namely  $\Tr(\tilde{\mathbf{Q}}^2)$ and $\Tr(\tilde{\mathbf{Q}}^3)$.  
Hence $\tilde{ f }_{Qb}$ is a polynomial 
in $\Tr(\tilde{\mathbf{Q}}^2)$ and $\Tr(\tilde{\mathbf{Q}}^3)$ 
with the only restriction on the expansion being that it must 
be stable against an unlimited growth of  $\tilde{\mathbf{Q}}$.  
As we will show the experimental data for $\tilde{S}$ in the 
nematic phase of CB7CB  fit  well to a model
where the expansion with respect to $\tilde{\mathbf{Q}}$ is 
taken at least  up to sixth
order terms. 
More specifically, in the absence of electric and magnetic fields,  
introducing $\tilde{I}_2=\Tr(\tilde{\mathbf{Q}}^2)$ and 
$\tilde{I}_3=\Tr(\tilde{\mathbf{Q}}^3)$, we take for the bulk free
energy density of the isotropic and the nematic phases
\begin{equation}
\tilde{f}_{Qb}=\tilde{f}_{Qb}[\tilde{I}_2,\tilde{I}_3]=
a_Q \tilde{I}_2-b \tilde{I}_3
+c \tilde{I}_2^2
   + d \tilde{I}_2 \tilde{I}_3 + e\left(\tilde{I}_2^3 -6 \tilde{I}_3^2\right)  + f \tilde{I}_3^2.
   \label{LdG_bulk}
\end{equation}
A  full account of phases, critical and tricritical points that this theory predicts 
is found in \cite{AllenderLonga&2008}.

The coefficients of the expansion (\ref{LdG_bulk}) generally depend on temperature  and other
thermodynamic variables, but in Landau theory the explicit 
temperature dependence is retained only in the bulk part, 
quadratic in $\tilde{\mathbf{Q}}$. In what follows, 
as a measure of temperature we choose the relative 
temperature distance, $\Delta t$, from the nematic-isotropic 
phase transition, defined through the relation
\begin{equation}\label{aQ}
a_Q=a_{0Q}\frac{(T-T^\ast)}{T_{NI}}=a_{0Q}\left(\frac{T-T_{NI}}{T_{NI}}+
\frac{T_{NI}-T^\ast}{T_{NI}}\right)=a_{0Q}(\Delta t+\Delta t_{NI}),
\end{equation}
where  $a_{0Q}>0$,  $T$ is the absolute temperature, $T_{NI}$ 
is the nematic-isotropic transition temperature, $T^\ast$ 
is the spinodal for a first-order
phase transition from the isotropic phase to the uniaxial nematic phase,
$\Delta t= (T-T_{NI})/{T_{NI}}\le 0$ and $\Delta t_{NI}=
(T_{NI}-T^\ast)/{T_{NI}}>0$ is the reduced temperature distance 
of nematic-isotropic transition temperature from  $T^\ast$. 
Additionally, 
$ b, c, d$, $e>0$ and $f>0$ are the temperature independent 
expansion coefficients. The last two
conditions for $e$ and $f$ guarantees that $\tilde{f}_{Qb}$
is stable against an unlimited growth of $\tilde{\mathbf{Q}}$ \cite{AllenderLonga&2008}.
The expansion, Eq.~(\ref{LdG_bulk}), generally  accounts  for the isotropic, 
uniaxial nematic and biaxial nematic phases \cite{AllenderLonga&2008,GRAMSBERGEN1986195}.

We should mention that the fourth order  
expansion, where $c>0$ and $d=e=f=0$ predicts that the $N_\text{TB}$ 
phase can be absolutely stable
within the family of one-dimensional modulated structures \cite{Pajak2018}, 
but the theory does not give a quantitative agreement with the data for
$\tilde{S}$ in the nematic phase of CB7CB unless unphysically 
large value of $\Delta t_{NI}$ is taken.
\subsection{Elastic free energy }
%%%%%%%%%%%%%%%%%%%%%%%
%
%%%% ELASTIC PART %%%%%
%
%%%%%%%%%%%%%%%%%%%%%%%
%
A spatial deformation of the alignment tensor $\tilde{\mathbf{Q}}$ 
in the nematic phase is measured by the  
elastic free energy density $\tilde{ f }_{Qel}$  of the 
Landau free energy expansion (\ref{eq:freeEnergy}).
For the description of elastic properties of nematic liquid crystals 
$\tilde{ f }_{Qel}$ usually is expanded into powers of 
$\tilde{\mathbf{Q}}$ and its first derivatives
$\partial\tilde{\mathbf{Q}}\equiv\partial \tilde{Q}_{ij}/
\partial \tilde{x}_k= \tilde{Q}_{ij,k}$, where only quadratic 
terms in derivatives of the order parameter field are retained,
in line with similar expansion for the director field, Eq.~(\ref{Oseen-Frank}).

This standard, the so called minimal-coupling Landau-de Gennes expansion for 
$\tilde{ f }_{Qel}$, comprises of only two elastic terms:  $\left[L^{(2)}_1\right]$ 
  $ = \tilde{Q}_{\alpha\beta,\gamma} \tilde{Q}_{\alpha\beta,\gamma} $
and $\left[L^{(2)}_2\right]$ 
$ = \tilde{Q}_{\alpha\beta,\beta} \tilde{Q}_{\alpha\gamma,\gamma}$.
Although again the  theory, Eq.~(\ref{eq:freeEnergy}), with $\tilde{ f }_{Qel}$
containing only these two elastic terms accounts for absolutely stable 
$N_\text{TB}$ among one-dimensional modulated structures \cite{Pajak2018} 
it is not sufficiently general to quantitatively reproduce, \emph{e.g.} 
elastic properties of bent-core systems 
in the parent nematic phase for it implies equality 
of  splay and bend Frank elastic constants, 
which so far is not an experimentally supported scenario with stable $N_\text{TB}$. 
Thus, we need to include higher-order elastic terms 
in the LdeG theory to account for experimentally observed 
elastic behaviour of bent-core mesogens.
A general form of the LdeG elastic free energy density 
has been studied by Longa \emph{et al.} in a series of papers 
\cite{LongaMonselesanTrebin1987,PhysRevA.39.2160,Longa&Trebin1990}. 
As it turns out the most important are third-order invariants 
of the form $\tilde{\mathbf{Q}}\,\partial \tilde{\mathbf{Q}}\,
\partial \tilde{\mathbf{Q}}$, given explicitly in Supplemental Material,
because they are the lowest order terms  
removing splay-bend degeneracy of the second-order theory
\cite{LongaMonselesanTrebin1987}.
But with quadratic and cubic terms alone the elastic free energy 
$\tilde{ f }_{Qel}$ is unbounded from below and, hence, 
cannot represent a correct theory of nematics.
To assure the nematic ground state is stable
against an unlimited growth of 
$\tilde{Q}_{\alpha\beta}$ and $\tilde{Q}_{\alpha\beta,\gamma}$
we need to add some fourth-order invariants 
\cite{LongaMonselesanTrebin1987}. 
In total, there are 22 deformation modes $\left[L^{(n)}_i\right]$  of $\tilde{\mathbf{Q}}$
up to the order $\tilde{\mathbf{Q}}\,\tilde{\mathbf{Q}}\,\partial\tilde{\mathbf{Q}}\,
\partial\tilde{\mathbf{Q}}$ (\emph{see} Supplemental Material for details).
The next step is to single out the relevant elastic terms 
$\left[L^{(n)}_i\right]$ that should enter the expansion $\tilde{ f }_{Qel}$. 
A~considerable reduction in the number of independent 
terms is obtained if we limit ourselves to a class of 
one-dimensional periodic structures  $\tilde{\mathbf{Q}}(\tilde{z} + 
\tilde{p}) = \tilde{\mathbf{Q}}(\tilde{z})$  \cite{supplemental,Pajak2018}.
 Then, the only relevant linearly independent  $[L]$-terms are 
\begin{itemize}
    \item $\partial \tilde{\mathbf{Q}}\,\partial \tilde{\mathbf{Q}}$
    terms: $\left[L^{(2)}_1\right]$, $\left[L^{(2)}_2\right]$
    \item $\tilde{\mathbf{Q}}\,\partial \tilde{\mathbf{Q}}\,\partial 
    \tilde{\mathbf{Q}}$ terms: $\left[L^{(3)}_2\right]$, $\left[L^{(3)}_3\right]$, $\left[L^{(3)}_4\right]$
    \item $\tilde{\mathbf{Q}} \tilde{\mathbf{Q}}\,\partial \tilde{\mathbf{Q}}\,
    \partial \tilde{\mathbf{Q}}$ terms: $\left[L^{(4)}_2\right]$, $\left[L^{(4)}_3\right]$, 
    $\left[L^{(4)}_5\right]$, $\left[L^{(4)}_6\right]$, $\left[L^{(4)}_7\right]$, $\left[L^{(4)}_{10}\right]$, $\left[L^{(4)}_{11}\right]$.  
\end{itemize}
As mentioned before  the most important are third order  terms 
$\tilde{\mathbf{Q}}\,\partial \tilde{\mathbf{Q}}\,\partial 
\tilde{\mathbf{Q}}$
linear in $\tilde{\QQ}$ and quadratic in $\partial \tilde{\mathbf{Q}}$, because they remove splay-bend degeneracy \cite{LongaMonselesanTrebin1987}. 
Hence, in what follows we will keep the third-order terms 
and add three stabilizing terms of the order 
$\tilde{\mathbf{Q}} \tilde{\mathbf{Q}}\,\partial \tilde{\mathbf{Q}}\,
\partial \tilde{\mathbf{Q}}$.
More specifically, for the elastic free energy $\tilde{f}_{Qel}$ we take a sum 
of quadratic terms in deformations of the form:
\begin{eqnarray}
    \tilde{f}_{Qel}&=& 
    L_1^{(2)}\tilde{Q}_{\alpha\beta,\gamma} \tilde{Q}_{\alpha\beta,\gamma} 
    + L_2^{(2)}\tilde{Q}_{\alpha\beta,\beta} \tilde{Q}_{\alpha\gamma,\gamma}+ 
     L_{14}^{(4)} \left( 
    \lambda_2 \tilde{Q}_{\mu\nu,\nu} + 
    \tilde{Q}_{\alpha\beta} \tilde{Q}_{\alpha\mu,\beta} \right)^2 + \nonumber\\
   && L_6^{(4)} \left( 
     \lambda_3 \tilde{Q}_{\beta\nu,\nu} + 
     \tilde{Q}_{\alpha\beta} \tilde{Q}_{\alpha\mu,\mu} 
     \right)^2
     + 
      L_7^{(4)} \left(  
      \lambda_4 \tilde{Q}_{\beta\mu,\nu} + 
      \tilde{Q}_{\alpha\beta} \tilde{Q}_{\alpha\mu,\nu} 
      \right)^2  
      \label{felFinal2}   \\
      &=&
    \left(L_1^{(2)}+ \lambda_4^2\,L_7^{(4)} \,\right) \left[L^{(2)}_1\right] + 
    \left(L_2^{(2)}+\lambda_2^2 L_{14}^{(4)} +\lambda_3^2 L_6^{(4)}\,\right) \left[L^{(2)}_2\right] +
    \nonumber \\
    && L_{2}^{(3)}  \left[L^{(3)}_2\right] + L_{14}^{(4)}  \left[L^{(4)}_{14}\right] +
    L_{3}^{(3)}  \left[L^{(3)}_3\right] + L_{6}^{(4)}  \left[L^{(4)}_{6}\right] +
    L_{4}^{(3)}  \left[L^{(3)}_4\right] + L_{7}^{(4)}  \left[L^{(4)}_{7}\right] 
    \label{felFinal}
\end{eqnarray}
where the coefficients $L^{(n)}_i$ denote temperature--independent elastic constants 
that couple with the invariant $\left[L^{(n)}_i\right]$ and where
$\lambda_2=L_2^{(3)}/\left(2 L_{14}^{(4)}\right)$,
$\lambda_3=L_3^{(3)}/\left(2 L_{6}^{(4)}\right)$, and
$\lambda_4=L_4^{(3)}/\left(2 L_{7}^{(4)}\right)$.
Use of the $\left[L_{14}^{(4)}\right]$ 
invariant, which is a linear combination of 
the invariants $\tilde{\mathbf{Q}} \tilde{\mathbf{Q}}\,\partial 
\tilde{\mathbf{Q}}\,\partial \tilde{\mathbf{Q}}$, allows to write the stability criteria
for $\tilde{f}_{Qel}$ in a simple form (\emph{see} Supplemental Material).
Indeed, the elastic free energy density $\tilde{f}_{Qel}$ is a sum of positive-definite terms
if
\begin{eqnarray}\label{StabilityCondition}
&& L_1^{(2)}>0, \hspace{0.5cm} L_1^{(2)}+\frac{2}{3} L_2^{(2)}>0, \nonumber\\
&&
  L_6^{(4)} > 0 \,\,\, \vee \,\,\, L_6^{(4)} = 0 \wedge L_3^{(3)} = 0, 
 \nonumber\\
 && L_7^{(4)} > 0 \,\,\, \vee \,\,\, L_7^{(4)} = 0 \wedge L_4^{(3)} = 0,
 \hspace{0.5cm} \nonumber\\
 && L_{14}^{(4)} > 0 \,\,\, \vee \,\,\, L_{14}^{(4)} = 0 \wedge L_2^{(3)} = 0.
\end{eqnarray}
The conditions (\ref{StabilityCondition})
are sufficient ones for $\tilde{f}_{Qel}$ to be positive definite $(\tilde{f}_{Qel}\ge 0)$.
For smooth tensor fields $\tilde{\QQ}$ 
the ground state of $\tilde{f}_{Qel}$  $(\tilde{f}_{Qel} = 0)$  corresponds 
to a constant, position-independent  $\tilde{\QQ}$, which represents 
an unperturbed uniaxial or biaxial nematic state. As we show in section 
\ref{estimates}  the elastic constants $L_m^{(n)}$ entering expansion 
(\ref{felFinal}) can all be estimated from the data for Frank elastic 
constants in the uniaxial nematic phase.  
To conclude, the free energy  (\ref{felFinal}) is a thermodynamically 
stable expansion of the Landau-de Gennes free energy in the 
local alignment tensor, complete up to third-order for deformations
realized in one spatial direction.

The elastic free energy (\ref{felFinal}) can still be written in a simpler form  
by further selecting terms that are relevant for UDS. Indeed, substitution of   
Eq.~(\ref{Q_tensor}) into expansion (\ref{felFinal}) induces  
extra relations between cubic and quartic  elastic invariants, namely
\begin{equation} \label{homogeneity_relations}
 \left[L^{(3)}_2\right] = -2 \left[L^{(3)}_3\right], \,\,\,\, 
 \left[L^{(4)}_{14}\right] = 4\left[L^{(4)}_6\right].
\end{equation}
Thus, in seeking for a relative stability of  UDS
two elastic terms in (\ref{felFinal}) are still redundant.
This redundancy becomes especially transparent in the 
parameterization where the elastic constants 
$L^{(3)}_2$, $L^{(3)}_3$, $L^{(4)}_6$ and $L^{(4)}_{14}$ are replaced by 
appropriate linear combinations of $L^{(3)}_7$, $L^{(3)}_8$, $L^{(4)}_{15}$ 
and $L^{(4)}_{16}$. They are given by
\begin{eqnarray} \label{newLs}
L^{(3)}_2 &=& L^{(3)}_8-L^{(3)}_7, \,\,\,\,\,\,\,\,
L^{(3)}_3 = 2\left( L^{(3)}_8 + L^{(3)}_7\right), \nonumber\\
L^{(4)}_6 &=& 4\left(L^{(4)}_{15}-L^{(4)}_{16}\right), \,\,\,\,\,\,\,\,
L^{(4)}_{14} =  L^{(4)}_{15} + L^{(4)}_{16}.
\end{eqnarray}
where, in addition, the inequality
$
L^{(4)}_{15} > \left| L^{(4)}_{16}\right|    
$
is required to fulfill  stability conditions (\ref{StabilityCondition}).
Substitution of (\ref{newLs}) into (\ref{felFinal}) now yields
\begin{eqnarray}
    \tilde{f}_{Qel}&=& 
    \left(L_1^{(2)}+ \lambda_4^2\,L_7^{(4)} \,\right) \left[L^{(2)}_1\right] + 
    \left(L_2^{(2)}+\lambda_2^2 L_{14}^{(4)} +\lambda_3^2 L_6^{(4)}\,\right) \left[L^{(2)}_2\right] +
    \nonumber \\
    && L_{7}^{(3)}  \left[L^{(3)}_7\right] + L_{15}^{(4)}  \left[L^{(4)}_{15}\right] +
    L_{8}^{(3)}  \left[L^{(3)}_8\right] + L_{16}^{(4)}  \left[L^{(4)}_{16}\right] +
    L_{4}^{(3)}  \left[L^{(3)}_4\right] + L_{7}^{(4)}  \left[L^{(4)}_{7}\right], 
    \label{felFinalFinal}
\end{eqnarray}
where 
\begin{eqnarray}\label{newElasticTerms}
     \left[L^{(3)}_7\right] &=& 2\left[L^{(3)}_3\right] - \left[L^{(3)}_2\right], \nonumber \\
  \,   \left[L^{(3)}_8\right]  &=& 2\left[L^{(3)}_3\right] + \left[L^{(3)}_2\right], \nonumber \\
  \,     \left[L^{(4)}_{15}\right] &=&  \left[L^{(4)}_{14}\right] + 4\left[L^{(4)}_{6}\right],  \nonumber \\
 \,     \left[L^{(4)}_{16}\right] &=& \left[L^{(4)}_{14}\right] - 4\left[L^{(4)}_{6}\right],
\end{eqnarray}
and where  $\left[L^{(3)}_8\right] $ and $\left[L^{(4)}_{16}\right]$ terms 
vanish for UDS, given by Eq.~(\ref{Q_tensor}).   
\subsection{Coupling with steric polarization}
According to the current understanding of  the formation 
of stable twist–bend nematic phase its  orientational order, 
being similar to that of smectic $C^\ast$ \cite{deGennesBook},
 should be accompanied with a long-range polar order of
 molecular bent cores
 \cite{GrecoLuckhurstFerrarini2014,Ferrarini2016,OsipovPajak2016,TomczykPajak2016,Tomczyk2020}. 
As already pointed out the other direct molecular
interactions, such as between electrostatic dipoles, are probably
less relevant for thermal stability of this phase.
Up to the leading order in $\tilde{\mathbf{P}}$ at least five  
extra terms must be included in $\tilde{f}_P$ and $\tilde{f}_{QP}$, 
Eq.~(\ref{eq:freeEnergy}). They read 
\cite{LongaMonselesanTrebin1987,LongaPajak2016,Pajak2018}
\begin{align}\label{fP}
\tilde{f}_P&=a_P\tilde{\mathbf{P}}^2+A_4(\tilde{\mathbf{P}}^2)^2
+b_P(\tilde{\nabla}\otimes\tilde{\mathbf{P}})^2
% +A_c(\tilde{\nabla}\cdot \tilde{\mathbf{P}})^2
,\\ \label{fQP}
\tilde{f}_{QP}&=-\varepsilon_P\tilde{\mathbf{P}}\cdot
(\tilde{\nabla}\cdot\tilde{\mathbf{Q}})-\Lambda_{QP}\tilde{P}_\alpha
\tilde{Q}_{\alpha\beta}\tilde{P}_\beta.
%\\
%f_{EQ}&=-\frac{1}{2}\varepsilon_0
%E_\alpha\Delta\varepsilon_{\alpha\beta}E_\beta-
%\varepsilon_{f}Q_{\alpha\beta}(E_{\alpha}\nabla_{\gamma}
%-E_{\gamma}\nabla_{\alpha})Q_{\beta\gamma}.
\end{align}
Here {$a_P=a_{0P}((T-T_P)/T_{NI})=a_{0P}(\Delta t+\Delta t_{NI,P})$, 
where $\Delta t_{NI,P}=(T_{NI}-T_P)/T_{NI}>0$}, $A_4>0$, $b_P>0$, 
% $A_c$,
$\varepsilon_P$ and $\Lambda_{QP}$ are further temperature-independent 
constitutive parameters of the model. Again, limitations for 
$A_4$ and  $b_P$ stem from stability requirement of a ground state
against unlimited fluctuations of $\tilde{\mathbf{P}}(\tilde{\mathbf{r}})$. 
The $\varepsilon_P$-term represents lowest-order flexopolarization contribution
while the $\Lambda_{QP}$-term is the direct coupling between the polarization 
field and the alignment tensor.  
The presently existing experimental data seem in line with 
this minimal coupling expansion for the (flexo)polarization
part of the free energy \cite{LongaPajak2016,Pajak2018}.
A full structure of the (flexo)polarization theory, along with some of its
general consequences, can be found in \cite{Longa&Trebin1990}.
\section{Reduced form of generalized Landau de-Gennes expansion}
For practical calculations 
 it is useful to reduce the number of model parameters by rewriting Eq.~(\ref{eq:freeEnergy}) 
in terms of reduced (dimensionless) quantities.
It reveals the redundancy of  four parameters in the expressions 
(\ref{LdG_bulk},\ref{felFinal},\ref{fP},\ref{fQP}) 
and allows to set them to one from the start 
\cite{Longa1986,LongaMonselesanTrebin1987,Longa&Trebin1990,LongaPajak2016}.
This reduction is a direct consequence  of the freedom to choose a scale 
for the free energy, $\tilde{F}=\Lambda_F\, F$,  for the fields $\tilde{\QQ}= 
\Lambda_Q\, \QQ$ and $\tilde{\PP}=\Lambda_P\, \PP$, 
and for the position vector $\tilde{\rr} = \Lambda_r\, \rr$, where $\Lambda_i$ 
are nonzero scaling parameters.
Taking this freedom into account  
we introduce the reduced quantities $F$ ($f_{tot}$), $\QQ $ (equivalently $S$, $x_0$, $r_1$, $r_2$),
$\PP $ (equivalently $p_1$, $v_0$), $\rr$, $\mathbf{k}$, 
$t_Q$, $\rho_{2,2}-\rho_{4,16}$, $ t_P$, 
$a_d$, $e_P$, $\lambda$,  $c_b$, $d_b$ and $e_b$ 
with the help of equations
\begin{eqnarray}\label{scaledVariables}
 \tilde{ F}  &=&  {\frac{b^2}{f}\, F}  \;\;\; 
 ( \tilde{ f}_{tot}  = \frac{b^2}{f}\, f_{tot}  ), \;\;\;
 { \tilde{\rr}   = \frac{\sqrt[6]{f}
   \sqrt{L^{(2)}_{1}}}{b^{2/3}} \rr, }  \;\;\;
{ \tilde{ \Q}   = \sqrt[3]{\frac{b}{f}}\,  \Q,}
\nonumber  \\ 
 \tilde{ \PP}  &=&  { \sqrt[3]{\frac{b}{f}}\,
\sqrt{\frac{ L^{(2)}_{1} }{b_P} }\, \PP,} \;\;\;  
{ a_Q = \sqrt[3]{\frac{b^{4} }{f}  }\, t_Q }, \;\;\;
{ a_P= \frac{b^{4/3}\, b_P\, t_P}{\sqrt[3]{f}\, L^{(2)}_{1}}}, 
 \nonumber\\
 L^{(2)}_{2}&=&L^{(2)}_{1} \rho _{2,2},\;\;\;
L^{(3)}_{i}= \sqrt[3]{\frac{f}{b}}\,
{ L^{(2)}_{1}\, \rho _{3,i}},\;\;\; 
% \label{aQ}
%
L^{(4)}_{i}= 
\sqrt[3]{\left(\frac{f}{b}\right)^2}\,
{ L^{(2)}_{1}\, \rho _{4,i}} 
\nonumber\\
\lambda_i &=& {L^{(2)}_{1}} l_i, \;\;\; 
l_2 = \frac{\rho _{3,2}}{2\, \rho _{4,14}},\;\;\;
l_3 = \frac{\rho _{3,3}}{2\, \rho _{4,6}},\;\;\;
l_4 = \frac{\rho _{3,4}}{2\, \rho _{4,7}},
\nonumber\\
%
%
% A_c&=& b_P\, a_c    ,\,\,\,\,
%
A_4&=&\frac{ \sqrt[3]{b^{2} f}\, a_d\,
b_P^2}{\left(L^{(2)}_{1}\right)^2}, \;\;\;
%
%\label{aP} \\
%
\varepsilon _P= 
\frac{b^{2/3} \sqrt{b_P}\, e_P}{\sqrt[6]{f}}, \;\;\;
\Lambda _{QP}=\frac{b \, b_P}{L^{(2)}_1}\, \lambda,
\nonumber\\
c&=& \sqrt[3]{b^{2} f}\, c_b,\;\;\;
%
% e=f e_b, \;\;\;
%
%\\ \label{labfield}
%
%\\ \label{labk}
%
d=\sqrt[3]{b f^{2}}\,  d_b,  \;\;\;
\tilde{\mathbf{k}} = \frac{ b^{2/3}  }{ \sqrt[6]{f}
   \sqrt{L^{(2)}_{1}} }
   \, \mathbf{k}.
\end{eqnarray}
The remaining quantities ($S$, $x_0$, $r_1$, $r_2$) and ($p_1$, $v_0$) are connected with their
tilted counterparts by the same relations as $\Q$ with $\tilde{ \Q}$ and $\PP$ with $\tilde{ \PP}$,
respectively. In addition, the definitions (\ref{newElasticTerms}) now become reduced to 
\begin{eqnarray}\label{newElasticTermsRed}
\rho _{3,2}&=&\rho _{3,8}-\rho _{3,7}, \,\,\,\,\,\, \rho _{3,3}=2 \left(\rho _{3,7}+\rho _{3,8}\right)
\nonumber\\
\rho _{4,6}&=&4 \left(\rho_{4,15}-\rho _{4,16}\right), \,\,\,\,\,\,\,
\rho _{4,14}=\rho _{4,15}+\rho _{4,16}.
\end{eqnarray}
Consequently, the generalized Landau-de Gennes free energy expansion
in terms of reduced variables ${\QQ}$ and ${\PP}$
reads  
\begin{equation}\label{eq:freeEnergyRed}  {
 { F}=\frac{1}{{ V}}\int_{{ V}}\, { f}_{tot}\, \mathrm{d}^{3}{ { {\rr}} }
    = \frac{1}{{V}}\int_{{ V}}\left({ f }_{b,Q} +
    { f }_{e,Q} + { f}_{P}+
    { f}_{QP} % + { f}_{E}
    \right)
    \mathrm{d}^{3}{{\rr}},}
\end{equation}
where 
\begin{align}
{f}_{Qb}&={f}_{Qb}[{I}_2,{I}_3]=
t_Q {I}_2 - {I}_3
+c_b {I}_2^2
   + d_b {I}_2 {I}_3 + e_b \left({I}_2^3 -6 {I}_3^2\right)  + {I}_3^2,
   \label{LdG_bulk_Red}
\\
\label{felFinalRed}
    {f}_{Qel}&= 
    \left(\,1+ \rho_{4,7}\, l_4^2\,\right){Q}_{\alpha\beta,\gamma}
    {Q}_{\alpha\beta,\gamma} + 
    \left(\rho_{2,2}+ \rho_{4,14}\, l_2^2 + \rho_{4,6}\, l_3^2\right)
    {Q}_{\alpha\beta,\beta} {Q}_{\alpha\gamma,\gamma}+ \nonumber \\
    & 
    (\,\rho_{3,8}-\rho_{3,7}\,)\,  {Q}_{\alpha\beta} {Q}_{\alpha\mu,\beta}
    {Q}_{\mu\nu,\nu}+
    2\,(\,\rho_{3,8}+\rho_{3,7}\,)\,  {Q}_{\alpha\beta} {Q}_{\alpha\mu,\mu}
    {Q}_{\beta\nu,\nu} + \nonumber \\
    &
    \rho_{3,4}  {Q}_{\alpha\beta} {Q}_{\alpha\mu,\nu}
    {Q}_{\beta\mu,\nu}+ 
    \rho_{4,7} {Q}_{\alpha\beta} {Q}_{\gamma\beta} 
    {Q}_{\alpha\mu,\nu} 
    {Q}_{\gamma\mu,\nu} + \nonumber \\
    &
    4\,(\rho_{4,15} - \rho_{4,16} )\,{Q}_{\alpha\beta} {Q}_{\gamma\beta} {Q}_{\alpha\mu,\mu}
    {Q}_{\gamma\nu,\nu}+
    (\rho_{4,15} + \rho_{4,16} ) {Q}_{\alpha\beta} {Q}_{\gamma\delta}
    {Q}_{\alpha\mu,\beta} 
    {Q}_{\gamma\mu,\delta},\\
\label{fP_Red}
{f}_P&=t_P{\mathbf{P}}^2+a_d\left({\mathbf{P}}^2\right)^2
+ ({\nabla}\otimes{\mathbf{P}})^2
%+a_c({\nabla}\cdot
%{\mathbf{P}})^2
,\\ \label{fQP_Red}
{f}_{QP}&=-e_P{\mathbf{P}}\cdot
({\nabla}\cdot{\mathbf{Q}})-\lambda{P}_\alpha
{Q}_{\alpha\beta}{P}_\beta,
%\\
%f_{EQ}&=-\frac{1}{2}\varepsilon_0
%E_\alpha\Delta\varepsilon_{\alpha\beta}E_\beta-\varepsilon_{f}Q_{\alpha\beta}
%(E_{\alpha}\nabla_{\gamma}-E_{\gamma}\nabla_{\alpha})Q_{\beta\gamma}.
\end{align}
and where $I_2=\Tr(\QQ^2)$ and $I_3=\Tr(\QQ^3)$. In this parameterization terms proportional
to $\rho_{3,8}$ and $\rho_{4,16}$ vanish for UDS.

The expansion (\ref{eq:freeEnergyRed}-\ref{fQP_Red})
is our GLdeG theory of modulated nematics.
If we limit ourselves to a family 
of periodic structures with  periodicity being 
developed in one spatial direction
the nonzero cubic and quartic couplings:
$\rho_{3,4}$, $\rho_{3,7}$,
$\rho_{4,7}$ and $\rho_{4,15}$ should admit the UDS states
as global minimizers. The remaining two couplings: $\rho_{3,8}$ and $\rho_{4,16}$ 
are solely responsible for one-dimensional, \emph{nonuniform} periodic distortions, 
which makes the corresponding elastic terms vanish for UDS. This means that 
depending on the choice of $\rho_{3,8}$ and $\rho_{4,16}$ we should be able to eliminate 
inhomogeneously deformed structures from the ground states of GLdeG, leaving only UDS. 
In the remaining part of this paper we are going to concentrate 
exclusively on this simpler case.
\section{Bifurcation scenarios for uniformly deformed structures}
Here we limit ourselves to the UDS structures  given 
in Table \ref{tab:structures} and determine bifurcation conditions 
for various symmetry breaking transitions.  Clearly, the isotropic-uniaxial 
nematic bifurcation temperature is given by $T^\ast$  (\ref{aQ}), which represents spinodal,
while the $I-N_\text{U}$ phase transition takes place at $T_{NI}$, slightly above $T^\ast$. Likewise,  
$T_P$ entering 
$a_P$, Eq.~(\ref{fP}), is transition temperature for a hypothetical phase 
transition from the isotropic to ferroelectric phase ($\PP\ne0$), in the absence 
of the nematic order ($\QQ = 0$). 
Both, $T^\ast$ and $T_P$ are examples of  bifurcation temperatures 
from less ordered phase (isotropic) - to more ordered one (nematic, ferroelectric). 
There are further bifurcations possible for UDS. 
Below, we give bifurcation conditions for all possible 
phase transitions between UDS given in Table~\ref{tab:structures}.  
The procedure is found in \cite{LongaPajak2016} and we only 
briefly sketch it below.
For given material parameters the equilibrium  amplitudes 
$y_i$ $\in$ $\{$ $r_1$, $r_2$, $p_1$, $x_0$, $v_0$ $\}$ 
are found from the minimization of the free energy $F$,  
Eq.~(\ref{felFinal2}), calculated for UDS 
(explicit formula for $F$ is listed in the Supplementary Material). 
They are solutions of a system of polynomial equations $\psi_i(t_Q,t_P,
\{y_\alpha\})\equiv\partial F / \partial y_i = 0$. 
In order to employ a bifurcation
analysis to $\{\psi_\alpha\}$  we expand
$y_i$, $t_Q$ and $t_P$  in an arbitrary parameter $\varepsilon$:
\begin{eqnarray} \label{epsilonExpansion}
y_i&=& y_{i,0} + \varepsilon y_{i,1} + \varepsilon^2 y_{i,2} + ... \nonumber\\
t_m&=& t_{m,0} + \varepsilon t_{m,1} + \varepsilon^2 t_{m,2} + ...\; \;\;\;\;\;\;\;\; m=Q,P,
\end{eqnarray}
where non-vanishing $y_{i,0}$ define the reference, 
higher symmetric phase. For example, if the reference 
state is the $N_\text{U}$ phase, only $y_{4,0} \equiv x_{0,0}$ is nonzero in Eqs.~(\ref{epsilonExpansion}).
By substituting Eqs.~(\ref{epsilonExpansion}) into $\{\psi_\alpha=0\}$ and letting  equations of the same order 
in $\varepsilon$ vanish we find equations for $y_{i,\alpha}$ and $t_{m},\; (m=Q,P)$. 
The leading terms, proportional to $\varepsilon^0$, are equations describing 
the high-symmetric reference state. Terms of the order $\varepsilon^1$ give 
conditions for  bifurcation to a low-symmetric phase. 
An equivalent approach would be to construct from $F$ the effective 
Landau expansion $\delta f(y_p)$ in a
primary order parameter $y_p$ of low-symmetric phase by 
systematically eliminating the remaining parameters $\{y_i\}$.
A detailed procedure is given in Ref. \cite{Longa1986}.
For example,  in case of the $N_\text{U}-N_\text{TB}$ phase transition 
we could take for the primary order parameter  either 
$y_1\equiv r_1$ or $y_3\equiv p_1$ with the final formulas being insensitive to the choice.
Before we start it proves convenient to introduce auxiliary variables 
$\kappa_1$ and $\kappa_2$: 
\begin{eqnarray}\label{kappaRelations}
\kappa _1 &=& 1 + l_4^2 \rho _{4,7}, \nonumber\\
\kappa _2  &=&  2\; \kappa _1+ \rho _{2,2}+l_2^2\; 
\rho _{4,14}+l_3^2\;
   \rho _{4,6},
\end{eqnarray}
and relative phases $\chi_1$, $\chi_2$:
\begin{eqnarray}\label{relativephases}
\chi _1 &=& \phi _1-\phi _p, \nonumber\\
\chi _2 &=& \phi _2-2 \phi _p, 
\end{eqnarray}
which simplify the free energy and consequently also bifurcation formulas. 
\subsection{\normalfont{$I-N_\text{TB}$}}
A bifurcation condition for a phase transition from 
the isotropic phase to the nematic twist-bend phase can be written down as
an equation connecting $t_P$ and $t_Q$:
\begin{equation}\label{landauITB}
a\equiv a(t_P,t_Q) = 2 \left({k}^2+{t_P}\right)-\frac{{e_P}^2 \text{k}^2 \sin
   ^2\left(\chi _1\right)}{2\; \kappa _2 {k}^2+4\; {t_Q}} = 0.   
\end{equation}
If we permit the $k-$vector and $\chi_1$ to vary, then 
the bifurcation temperature $t_{P,I-TB}$ from the isotropic to nematic 
twist-bend phases will be the
maximal $t_P$ fulfilling the condition (\ref{landauITB}). 
Solving Eq.~(\ref{landauITB}) for $t_P$ and maximizing with respect to 
$k$ and $\chi$  gives the bifurcation values for $k$, $\sin(\chi_1)$ and $t_P$.
They read
\begin{eqnarray}\label{bif-I-TB}
k^2_{I-TB}&=&\max \left(0,\frac{\sqrt{{t_Q}} \left| {e_P}\right| }{\sqrt{2}
   \left| \kappa _2\right| }-\frac{2 {t_Q}}{\kappa _2}\right), \,\,\,\,\, \textrm{for}\,\,\, 
   t_Q>t_{NI}>0, \,\,\,\, \kappa_2>0  
   \nonumber\\
\sin(\chi_{1,I-TB})^2&=&1 \nonumber \\
t_{P,I-TB}&=& 
\left\{
\begin{array}{cc}
0 &  \,\,\,\,\, \textrm{for}\,\,\, k_{I-TB}=0 \\
\frac{{e_P}^2+8
   {t_Q} -4 \sqrt{2} \sqrt{{t_Q}} \left| {e_P}\right| }{4 \kappa _2} &  \,\,\,\,\, \textrm{for}\,\,\, k_{I-TB}>0 
\end{array}
\right. 
\end{eqnarray}
where $t_{NI}>0$ is the isotropic-uniaxial nematic transition temperature. 
Using formalism of Ref. \cite{Longa1986} one can also show that
the $a$-term, Eq.~(\ref{landauITB}), is actually the leading coefficient
in the Landau expansion of the free energy of the $N_\text{TB}$ 
phase about the reference $I$ phase in the primary order parameter $p_1$:
\begin{equation}\label{landauinp1}
\Delta f_{I-TB} = \frac{1}{2!} a\, p_1^2 + \frac{1}{4!} b\, p_1^4 +
\frac{1}{6!} c\, p_1^6 + ...    
\end{equation}
Generally,  nonzero value $k_{I-TB}$ of 
the $k$-vector at the bifurcation point ($e_P^2>8t_q$) indicates 
that the $I-N_\text{TB}$ phase transition is, at least weakly,
first order. It can be classified as an example 
of a weak crystallization introduced by Kats {\emph et al.} \cite{KATS19931}, with 
fluctuations that should be observed near the  $k = k_{I-TB}$ sphere.
Interestingly, for $k=k_{I-TB}=0$  a direct inspection of higher-order terms of  
the expansion (\ref{landauinp1}) shows that $b=24\, a_d-\frac{\lambda^2}{t_Q}$ can change 
sign for sufficiently large $\lambda$. That is, for $c>0$ the $I-N_\text{TB}$ transition 
can be first order ($a=0$, $b<0$),
second order ($a=0$, $b>0$), or  tricritical ($a=0$, $b=0$).
\subsection{\normalfont{$N_\text{U}-N_\text{TB}$}}
A bifurcation condition from $N_\text{U}$ to $N_\text{TB}$, 
expressed in terms of an equation connecting $t_P$ and $x_0$ reads:
\begin{eqnarray}\label{landauNUNTB}
a\equiv a(t_P,x_0)=
2 \left({k}^2+{t_P}+\frac{\lambda \; x_0}{\sqrt{6}}\right)
-\frac{3\; {e_P}^2 \sin ^2\left(\chi _1\right)}
 {6\; \kappa _2+x_0 \left(\sqrt{6}\; \kappa_3+\kappa _4 x_0\right)} =0
\end{eqnarray}
%
%
%\begin{equation}
% t_P= -k^2 -\frac{\lambda  x_0}{\sqrt{6}} + \frac{3 e_P^2 \sin ^2
%\left(\chi _1\right)}{2 \left(6 \kappa _2
% +\sqrt{6} \kappa _3 x_0
% +\kappa _4  x_0^2\right),}   
%\end{equation}
%
where  $\kappa _3=\rho _{3,4}-4 \rho _{3,7}$, $\kappa _4=5 \rho _{4,7}+
8 \rho _{4,15}$ 
and where $x_0$ is the nematic order parameter calculated from the minimization 
of $f_{Qb}$ in the uniaxial nematic phase. For fixed $t_Q$ ($x_0$) 
the  bifurcation temperature corresponds to the maximal $t_P$ fulfilling 
the condition of $a=0$, Eq.~(\ref{landauNUNTB}), 
where maximum is taken over $k$ and $\chi_1$.
It yields
\begin{equation}\label{bif-N-TB}
 t_{P,N-TB}= -\frac{\lambda  x_0}{\sqrt{6}} + \frac{3 e_P^2 }{2 \left(6 \kappa _2
 +\sqrt{6} \kappa _3 x_0
 +\kappa _4  x_0^2\right)}
\end{equation}
for $k=0$ and $\sin ^2\left(\chi _1\right)=1$.
As previously for the $I-N_\text{TB}$ phase transition, 
the $a$-term, Eq.~(\ref{landauNUNTB}), is the leading coefficient
in Landau expansion of the free energy of the $N_\text{TB}$ phase about 
the reference $N_\text{U}$ phase in the primary order parameter $p_1$ \cite{Longa1986}: 
$\Delta f_{N-TB} = \frac{1}{2!} a p_1^2 + \frac{1}{4!} b p_1^4 +
\frac{1}{6!} c p_1^6 + ... $ .
A direct inspection of this expansion shows 
that the $N_\text{U}-N_\text{TB}$ transition can  be first order ($a=0$, $b<0$, $c>0$),
second order ($a=0$, $b>0$, $c>0$), or tricritical ($a=0$, $b=0$, $c>0$).
Our analysis in the next section shows that for the case of CB7CB 
the predicted $N_\text{U}-N_\text{TB}$ transition is weakly first-order.
The  tricritical conditions for  $I-N_\text{TB}$ 
and $N_\text{U}-N_\text{TB}$ phase transitions
will be studied in detail elsewhere.
\subsection{Bifurcations to globally polar phases}
In a similar way  we can derive the  bifurcation condition for 
phase transitions from $I$, $N_\text{U}$, $N_\text{TB}$ to corresponding globally polar 
structures listed in Table \ref{tab:structures}. It reads
\begin{eqnarray}\label{landau-polar-structures}
%
% a  &=&2\; {t_P}=0, \hspace{4cm}
% \text{for} \,\,\,\,\, I-N_{U,p} \\
%
% a  &=&2\; {t_P}-2 \sqrt{\frac{2}{3}} \lambda  x_0 =0, 
%\hspace{2cm} \text{for} \,\,\,\,\,  N_U-N_{U,p}
%
%\\
%
a  &=&2\; {t_P}-2 \sqrt{\frac{2}{3}} \lambda  x_0 + 4 {a_d}\; p_1^2
=0, %\,\,\,\,\,\,\,\, \text{for} \,\,\,\,\, N_{TB}-N_{TB,p}
\end{eqnarray}
where $x_0=p_1=0$ for $ I-N_{\text{U},p}$ bifurcation, $p_1=0$ for $N_\text{U}-N_{\text{U},p}$ bifurcation, and where
both $x_0$ and  $p_1$ are non-zero when bifurcation takes place from $N_\text{TB}$ to $N_{\text{TB},p}$.
Now the parameter $a$ is the leading coefficient of Landau expansion  $\Delta f_{p} = \frac{1}{2!} a v_0^2 +
\frac{1}{4!} b v_0^4 +
\frac{1}{6!} c v_0^6 + ... $
in $v_0$- the primary order
parameter for phase transitions to polar structures. Given the form of 
the expansion for $f_P$, Eq.~(\ref{fP}), the tricritical point
can only be found for the $N_\text{TB}- N_{\text{TB},p}$ phase transition.
\section{Estimates of model parameters from experimental data for CB7CB  \label{estimates}}
Before we explore relative  stability of the nematic phases 
given in Table \ref{tab:structures} we estimate some of the material  
parameters entering the expansion (\ref{eq:freeEnergy}) 
from experimental data  in the uniaxial 
nematic phase.  This will allow us to study properties of $N_\text{TB}$ 
with only a few adjustable parameters.
Indeed,  nearly all of the parameters of the purely nematic parts: 
the bulk $\tilde{f}_{Qb}$  and the elastic $\tilde{f}_{Qel}$ can be correlated with  
the existing data in the uniaxial nematic phase. 
The $N_\text{U}$ phase  usually appears stable at higher 
temperatures and $N_\text{U}-N_\text{TB}$ phase transition is 
observed as temperature is lowered.  

 The very first compound shown to exhibit stable nematic and twist-bend nematic 
phase was the liquid crystal dimer 
1$^{\prime\prime}$,7$^{\prime\prime}$-bis(4-cyanobiphenyl-4$^{\prime}$-yl)heptane, 
abbreviated as CB7CB \cite{Cestari2011,Borshch2013,Chen2013}. This compound is constituted of two 4-cyanobiphenyls (CB) linked by an alkylene spacer (C$_7$H$_{14}$). Currently, it is one of the 
best studied examples.  In particular,  Babakhanova \emph{et al.} 
\cite{Babakhanova2017} have carried out a series of experiments
for this mesogen
in the uniaxial nematic phase. They determined 
the temperature variation of the nematic order parameter 
$\tilde{S}$ (\emph{see} Eq.~(\ref{alignmentTensor})), the temperature variation 
of the Frank elastic constants $K_{ii}$ $(i=1,2,3)$, 
the nematic-isotropic transition temperature $T_{NI}$, 
the nematic twist-bend-nematic transition temperature $T_{NN_{\text{TB}}}$.
We use the data presented in Table \ref{tab:CB7CBdata} to estimate some of the
parameters of the extended LdeG theory.
\subsection{Bulk part}
\begin{figure}[h!]
    \centering
    \includegraphics[width=.71\textwidth]{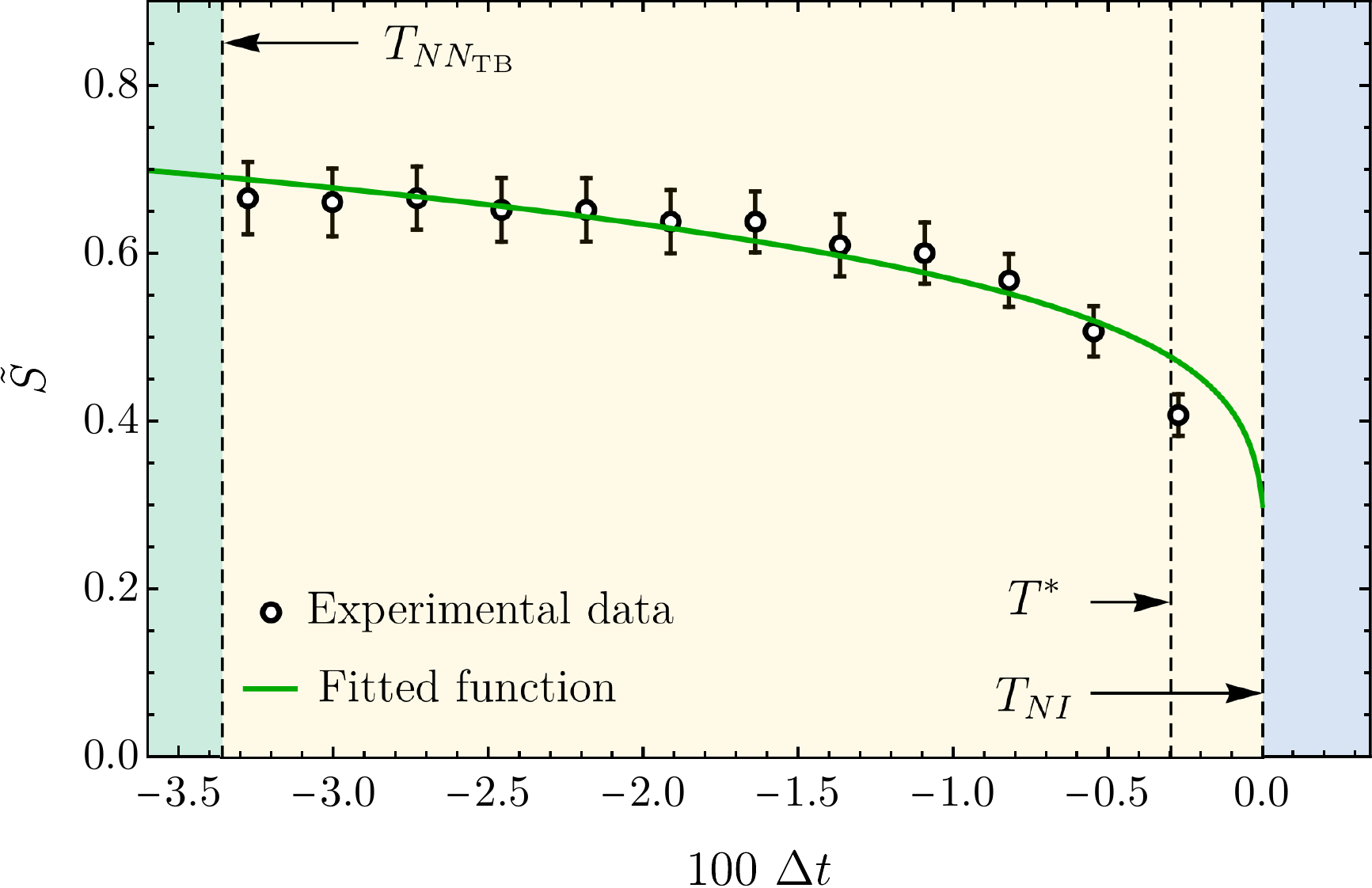}
    \caption{Experimental data from \cite{Babakhanova2017} 
    representing the temperature dependence of $\tilde{S}$ in 
    the uniaxial nematic phase of CB7CB. Green line illustrates 
    the effect of fitting predictions of theory 
    (\ref{LdG_bulk_S_parameter})  
    to the data. $T^\ast$ represents the maximal supercooling 
    temperature of the isotropic phase.
    Our fit is carried out by taking as \emph{ansatz} the experimentally 
known value of $T_{NI}$ and $T^\ast$. Then $\tilde{S}_{NI}$ for CB7CB is estimated
from our fitted function. If we compromise the agreement of $T_{NI}$ and $T^\ast$ 
with experimental data a better fit can be obtained for $\tilde{S}$ close 
to the transition temperature without affecting one in the vicinity of $N_\text{U}-N_\text{TB}$.}
    \label{fig:1}
\end{figure}

Under the assumption that $\tilde{\mathbf{Q}}$ is uniaxial and 
positionally independent (Eq.~\ref{alignmentTensor} with 
$\nn(\tilde{\rr})=\textrm{const.}$), 
the order parameter $\tilde{S}$ can be determined 
from the minimum of the free-energy 
density $\tilde{f}_{Qb}$, Eq.~\ref{LdG_bulk}, 
which becomes reduced to that of the uniaxial nematic phase
\begin{equation}
\tilde{f}_{Qb}=\frac{2}{3}a_{0Q}(\Delta t+\Delta t_{NI}) 
\tilde{S}^2-\frac{2}{9}b \tilde{S}^3+\frac{4}{9}c \tilde{S}^4
   + \frac{4}{27}d \tilde{S}^5 + {\frac{4}{81} f} \tilde{S}^6.
   \label{LdG_bulk_S_parameter}
\end{equation}
Now, from the necessary condition for minimum, 
$\partial\tilde{f}_{Qb}/\partial \tilde{S}=0$, solved for $T(\tilde{S})$, 
we determine the ratios $b/a_0 =\tilde{b}$, $c/a_0= \tilde{c}$, $d/a_0=\tilde{d}$ 
and {$f/a_0=\tilde{f}$} by fitting $\{\tilde{S}, 
T(\tilde{S})\}$ to the experimental data  of 
Babakhanova \emph{et al.} \cite{Babakhanova2017}.  
Independently, the scaling factor $a_{0}$ can be estimated 
from the latent heat per mole
$\Delta H_{NI}= T_{NI} \left(\partial \tilde{f}_{Qb}/\partial T \right)
_{T\to T_{NI}^{-}}$ for the  nematic–isotropic phase transition.
It reads
\begin{equation}
    a_0=\frac{a_{0Q}}{T_{NI}}=\frac{3 \Delta H_{NI} (2 \tilde{S}_{NI} (24 
    \tilde{c}+\tilde{S}_{NI} (15 \tilde{d}+8 \tilde{f}
   \tilde{S}_{NI}))-9 \tilde{b})}{2 \tilde{S}_{NI}T_{NI}\left(9 \tilde{b}
   \tilde{S}_{NI}+10 \tilde{d} \tilde{S}_{NI}^3+8 \tilde{f} \tilde{S}_{NI}^4-36 T_{NI}+36
   T^\ast\right)},
\end{equation}
where $\tilde{S}_{NI}$ is the nematic order parameter 
at the uniaxial nematic--isotropic phase transition. 
{ For overall consistency, $a_{0}$  
has been onward multiplied by 
$\rho_{C}=\rho/M_{w}\approx 2.22*10^3$\,\si{mol/m^3}, 
which is the ratio of the mass density $\big(\rho\approx 10^{3}$\,
\si{kg/m^3}$\big)$ to the relative molecular weight of 
CB7CB ($M_{w}\approx0.45$\,\si{kg/mol}) yielding the value $6.88\cdot 10^4$ J m$^{-3}$ K$^{-1}$. Thanks to this operation it was possible to express all expansion coefficients 
in units \si{J/m^{-3}}.}
Fig. \ref{fig:1} depicts results of fitting to experimental data, whereas numerical
values of the parameters are gathered in Table \ref{tab:fitbulkelastic}. Please observe that
the expansion parameter $e$ couples to a purely biaxial part and therefore it cannot be estimated
from the data in $N_\text{U}$. 
%

%%%%%%%%%%%%%%%%%%%%%%%%%%%%%%%%%%%%%%%%%%%%%%%%%
\begingroup
\renewcommand*{\arraystretch}{1.5}
\begin{table}[htp!]
\caption{Basic experimental data for CB7CB used to estimate some of the parameters of the extended LdeG thoery, along with other crucial data based on aforementioned estimates.}
\begin{tabular}{c|c|c|c|c}
\hline
\multicolumn{5}{c}{\textbf{Nematic}} \\ \hline
 Description      & Parameter & Value   & Unit   & Source   \\\hline
  \begin{minipage}[c]{0.4\columnwidth}Temperature of nematic-isotropic phase transition\end{minipage}& $T_{NI}$ & 387.15   &    \si{K}   & \cite{Babakhanova2017} 
                  \\\hline 
 \begin{minipage}[c]{0.4\columnwidth}Supercooling temperature of the isotropic phase\end{minipage}                  &$T^\ast$&386&\si{K}&\begin{minipage}[c]{0.2\columnwidth}acquired from $\tilde{S}(T)$\end{minipage}\\\hline
  Enthalpy                & $\Delta H_{NI}$ & 0.72   &    \si{kJ/mol}   
  & \cite{Paterson2017} \\\hline
  Order parameter at $T_{NI}$& $\tilde{S}_{NI}$&0.3&-&\begin{minipage}[c]{0.2\columnwidth}acquired from $\tilde{S}(T=T_{NI})$\end{minipage}\\\hline
\multicolumn{5}{c}{\textbf{Twist-bend nematic}} \\ \hline
 Description      & Parameter & Value    & Unit   & Source   \\\hline
\begin{minipage}[c]{0.4\columnwidth}Temperature of nematic-twist-bend nematic phase transition\end{minipage} & $T_{N N_\text{TB}}$ & 374.15 & \si{K} & \cite{Babakhanova2017}\\\hline
Enthalpy & $\Delta H_{N N_\text{TB}}$ & 0.82   &    \si{kJ/mol}   
& \cite{Paterson2017} \\\hline
\multicolumn{5}{m{15cm}}{\small{Twist-bend nematic phase can be supercooled 
to about 304.15 \si{K} \cite{Cestari2011} and then there  
is a~glass  transition  at  approximately 277.15 \si{K} \cite{Lopez2012}.}} \\ \hline
\end{tabular}
\label{tab:CB7CBdata}
\end{table}
\endgroup

%
%
%
%
%%%%%%%%%%%%%%%%%%%%%%%%%%%%%%%%%%%
%%%%Experimental Data%%%%%%%%%%%%%%
%%%%%%%%%%%%%%%%%%%%%%%%%%%%%%%%%%%

%
%
%
%
\subsection{Flexopolarization renormalized elasticity of uniaxial nematics}
It is important to realize that although (flexo)polarization terms (\ref{fP},\ref{fQP}) vanish 
in the uniaxial nematic phase any local deformation of the alignment tensor
induces deformation of $\tilde{\mathbf{P}}$ due to the flexopolarization coupling
$\varepsilon_P \ne0$. Such deformations  effectively renormalize 
elastic constants $L^{(n)}_m$ in ordinary nematic phases. The effect cannot be neglected
if we intend to estimate $L^{(n)}_m$ from experimental data. A mathematical procedure
of taking into account such deformations in the $N_\text{U}$ phase is to minimize 
the free energy Eq.~(\ref{eq:freeEnergy}) 
over Fourier modes of the polarization field for given, fixed Fourier modes
of the alignment tensor. Assuming that 
deformation 
$\tilde{\mathbf{Q}}(\tilde{\mathbf{r}})$ is small and slowly varying we obtain
with this procedure the $\tilde{\mathbf{Q}}$-induced deformation  
of $\tilde{\mathbf{P}}(\tilde{\mathbf{r}})$ expressed in terms of Fourier modes which, when  
transformed back to the real space take the form of a series in $Q_{\alpha\mu }$, 
$Q_{\alpha\mu,\mu }$ and in higher-order derivatives of $Q_{\alpha\mu }$. 
The relevant terms are
\begin{equation}{\label{P-deformations}}
 \tilde{P}_\alpha(\tilde{\mathbf{r}}) =
 \frac{\varepsilon _P}{2 a_P}  \tilde{Q}_{\alpha\mu,\mu }
    + \frac{\varepsilon _P \Lambda _{\text{QP}} }{2 a_P^2}
   \tilde{Q}_{\alpha \beta } \tilde{Q}_{\beta\mu,\mu }
   + \frac{\varepsilon _P \Lambda _{\text{QP}}^2 }{2 a_P^3}
   \tilde{Q}_{\alpha \lambda } \tilde{Q}_{\lambda \beta  }
   \tilde{Q}_{\beta\mu,\mu } + \dots
\end{equation}
Substituting Eq.~(\ref{P-deformations})  back to $\tilde{f}_P$ and $\tilde{f}_{QP}$ 
we obtain effective elastic contributions expressed in terms of only 
$\tilde{Q}_{\alpha\beta}$ and  $\tilde{Q}_{\gamma\mu,\mu }$.
When added to $ \tilde{f}_{Qel}$ they give an effective elastic free energy of uniaxial 
and biaxial nematics with $ L^{(n)}_m$ being replaced by 
$L^{(n)}_{m,eff}$, where relevant $ L^{(n)}_m$'s are
\begin{eqnarray}\label{P-renormalized-L}
 L^{(2)}_2 & \rightarrow & L^{(2)}_{2,eff} = L^{(2)}_2 -
 \frac{\varepsilon _P^2}{4\, a_P}, \nonumber \\
L^{(3)}_3 & \rightarrow & L^{(3)}_{3,eff} = L^{(3)}_3 -
 \frac{\Lambda _{\text{QP}} \,\varepsilon _P^2}{4\, a_P^2}, \nonumber \\
L^{(4)}_6 & \rightarrow & L^{(4)}_{6,eff} = L^{(4)}_6 -
\frac{\Lambda _{\text{QP}}^2\, \varepsilon
   _P^2}{4\, a_P^3}.
\end{eqnarray}
An important physical distinction between the bare constant $L^{(n)}_m$ 
and the renormalized constant $L^{(n)}_{m,eff}$ is of the same sort as one between 
renormalized and bare Frank elastic constants, as discussed by 
J\'akli, Lavrentovich and Selinger
\cite{RevModPhys.90.045004}: $L^{(n)}_m$ gives the
energy cost of  $\tilde{Q}_{\alpha\beta,\gamma}$ deformations if we 
constrain $\tilde{P}_\alpha=0$ during the deformation, while $L^{(n)}_{m,eff}$
relaxes to its optimum value during the the deformation.  
Assuming that major contribution to flexopolarization 
is of entropic, excluded volume type, any realistic experiment 
to measure elastic constants should not put constraints 
on the  polar field $\tilde{\PP}$, but rather allows it to relax. 
In this case, which is analysed here, $L^{(n)}_{m,eff}$ are the relevant contributions
to the elastic constants  in Eq.~(\ref{felFinal}).
\subsection{Elastic part}
\begin{figure}[htp!]
    \centering
    \includegraphics[width=0.6\textwidth]{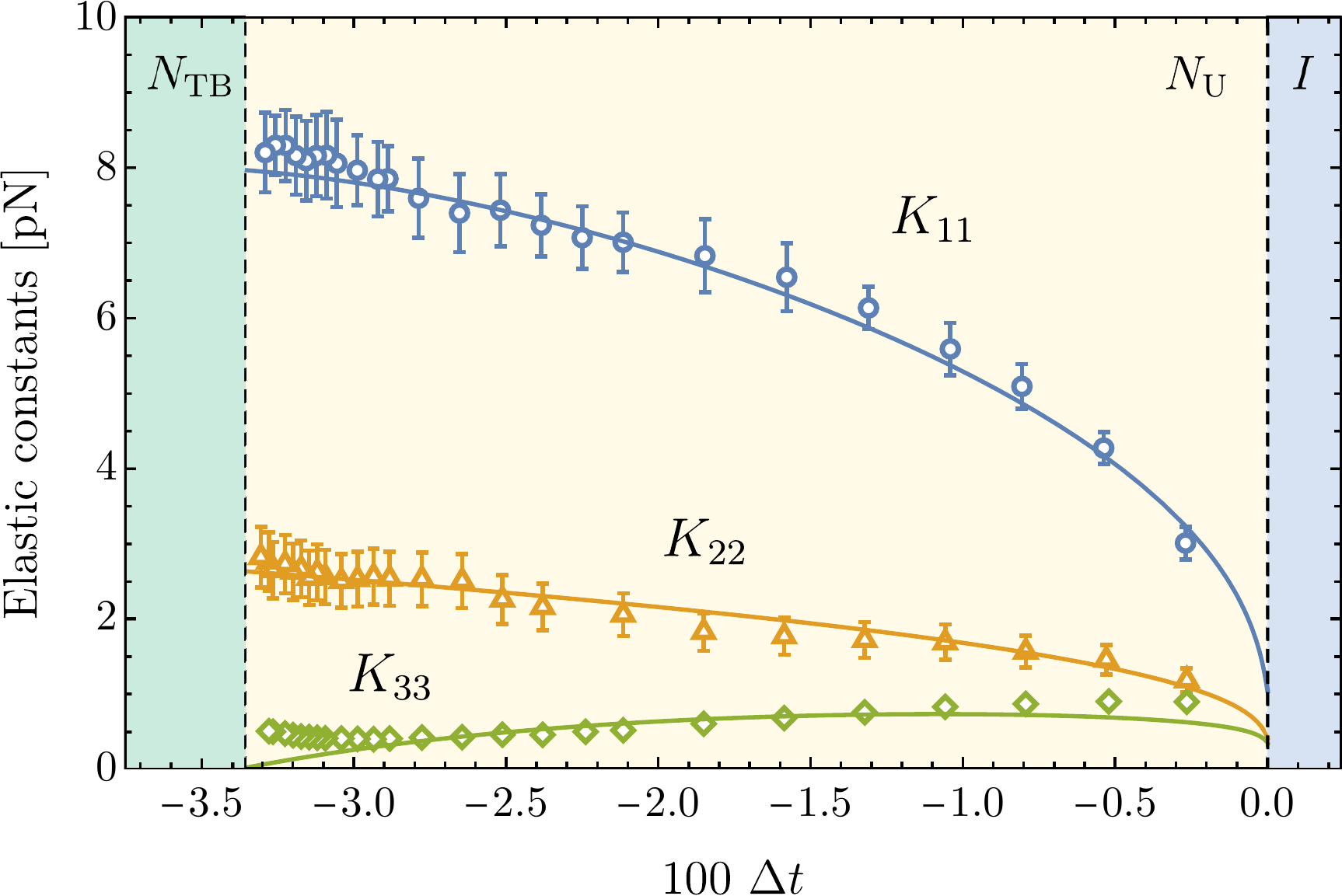}
    \caption{Temperature dependence of elastic constants acquired from \cite{Babakhanova2017}. 
    Continuous lines depict adopted approach for elastic constants within the model.
    Note that the model cannot explain a slight increase of $K_{33}$ in the vicinity
    of the $N_\text{U}-N_\text{TB}$ phase transition.}
    \label{elasticconstantsfit}
\end{figure}

In the hydrodynamic limit 
where spatial dependence of $\tilde{S}$ is disregarded and $\tilde{\mathbf{Q}}$ is given by 
(\ref{alignmentTensor}) the elastic free energy $\tilde{ f }_{Qel}$ 
turns into the Oseen-Frank free energy density of the director 
field $\mathbf{n}(\tilde{\rr})$,
 Eq.~(\ref{Oseen-Frank}), with  
$K_{11}$, $K_{22}$ and $K_{33}$ being polynomials in $\tilde{S}$ \cite{LongaMonselesanTrebin1987}
\begin{equation} \label{KseriesInS}
    K_{ii}= K_{ii}^{(2)}\tilde{S}^2+K_{ii}^{(3)}
    \tilde{S}^3+K_{ii}^{(4)}
    \tilde{S}^4 \quad\quad(i=1,2,3).
\end{equation}
The coefficients $K_{ii}^{(n)}$ are functions of $L^{(n)}_j$
\cite{LongaMonselesanTrebin1987,PhysRevA.39.2160} fulfilling
the splay-bend degeneracy relation in the second order:
$K_{11}^{(2)}=K_{33}^{(2)}$.  
For completeness, they are given in Supplemental Material.
As it turns out $K_{ii}^{(n)}$ 
with $n=2,3,4$ along with flexopolarization renormalization (\ref{P-renormalized-L})
are sufficient to obtain a good fit of  Eq.~(\ref{KseriesInS}) 
to  experimentally observed $K_{ii}$
for CB7CB \cite{Babakhanova2017}. 
Importantly, they also provide an estimate for the (flexo)polar couplings 
$\varepsilon_P$ and $\Lambda_{QP}$.  
In finding $K_{ii}^{(n)}$ we use the  
$\tilde{S}(T)$ fit obtained from the analysis of the bulk free energy 
in Section (\ref{BNP}), which is a prerequisite to have a consistent theory 
of the uniaxial nematic phase for this compound. 
Results of fitting are gathered in Table \ref{tab:fitbulkelastic}.
Finally, as the number of relevant couplings 
$L_\alpha^{(n)}$ ($n=3,4$), Eq.~(\ref{felFinal}), equals that of $K_{ii}^{(n)}$  
we can correlate $L_\alpha^{(n)}$ with $K_{ii}^{(n)}$ using the results 
of Supplemental Material and of Ref. \cite{LongaMonselesanTrebin1987}. It yields  
\begin{eqnarray} \label{LvsK}
L_1^{(2)}+ \lambda_4^2\,L_7^{(4)} \, &=& \frac{1}{4} K_{22}^{(2)}, \nonumber \\
L_2^{(2)}+\lambda_2^2 L_{14}^{(4)} +\lambda_3^2 L_6^{(4)}\,
- \frac{\varepsilon _P^2}{4\, a_P}  
&=& \frac{1}{2}\left( K_{11}^{(2)}- K_{22}^{(2)} \right),
\nonumber \\
L_2^{(3)} &=& \frac{1}{2}\left( K_{11}^{(3)}- 3 K_{22}^{(3)}
+ 2 K_{33}^{(3)} \right),
\nonumber \\
L_3^{(3)} -
 \frac{\Lambda _{\text{QP}} \,\varepsilon _P^2}{4\, a_P^2} 
 &=& \frac{1}{2}\left( 2 K_{11}^{(3)}- 3 K_{22}^{(3)}
+  K_{33}^{(3)} \right),
\nonumber \\
L_4^{(3)} &=& \frac{3}{2} K_{22}^{(3)},  \\
L_6^{(4)}-
\frac{\Lambda _{\text{QP}}^2\, \varepsilon
   _P^2}{4\, a_P^3} &=& \frac{3}{10}\left( 4 K_{11}^{(4)}- 3 K_{22}^{(4)}
-  K_{33}^{(4)} \right),
\nonumber \\
L_7^{(4)} &=& \frac{9}{10} K_{22}^{(4)}, \nonumber \\
L_{14}^{(4)} &=& -\frac{3}{10}\left( K_{11}^{(4)}+ 3 K_{22}^{(4)}
- 4 K_{33}^{(4)} \right).\nonumber
\end{eqnarray}
Results of fitting of $L_\alpha^{(n)}$, Eq.~(\ref{felFinal}),
obeying stability ansatz (\ref{StabilityCondition})
to experimental data for CB7CB are given in Table \ref{tab:fitbulkelastic}.
Quality of fit is displayed in Fig. \ref{elasticconstantsfit}. 
\begingroup
\renewcommand*{\arraystretch}{1}
\begin{table}[htp!]
\caption{Values of fitted coefficients of the bulk (\ref{LdG_bulk_S_parameter}) and elastic constants (\ref{KseriesInS}) expansions, along with ones resulting from the flexopolarization renormalization. Additionally, according to (\ref{felFinal}) and (\ref{LvsK}) are provided $K_{ii}^{(n)}$ elastic constants.}
\begin{tabular}{cc|cc|cc}
\colrule
Coefficient & Value $\big[\times10^7\,$\si{J/m^3}$\big]$ &Coefficient & Value [pN] & Coefficient & Value [pN] \\
\colrule
$a_{0Q}$& 2.66 & $K^{(2)}_{11}=K^{(2)}_{33}$ & 3.54 & $L_1^{(2)}$ & 0.93
 \\ 
$b$ & 0.27 & $K^{(2)}_{22}$ & 3.72  & $L_2^{(2)}$ & $-0.0045$\\ 
$c$&0.60&$K^{(3)}_{11}$ & 11.08 & $L_2^{(3)}$ & $-11.16$  \\ 
$d$&-3.80&$K^{(3)}_{22}$ & 0.60 & $L_3^{(3)}$ & 2.27\\ 
$f$&9.64&$K^{(3)}_{33}$ & $-15.80$  & $L_4^{(3)}$ & 0.90\\ 
&&$K^{(4)}_{11}$ & 9.93 & $L_6^{(4)}$ & 6.29 \\ 
&&$K^{(4)}_{22}$ & 1.77  & $L_7^{(4)}$ & 1.59\\ 
&&$K^{(4)}_{33}$ & 13.45 & $L_{14}^{(4)}$ & 11.56\\ 
&&&&$\frac{\varepsilon^2_{P}}{4a_{P}}$&0.08\\
&&&&$\frac{\Lambda_{\text{QP}}\varepsilon^2_{P}}{4a^2_{P}}$&0.00013\\
\colrule
\multicolumn{6}{c}{$T_P=362$ K}\\
\colrule
\end{tabular}
\label{tab:fitbulkelastic}
\end{table}
\endgroup
\newpage

\section{Predictions for nematic twist-bend phase of CB7CB-like compounds}

Within this section, we explore the relative stability of the UDS structures, listed in 
Table~\ref{tab:structures}, for the model Eq.~(\ref{eq:freeEnergyRed}) with parameters (estimated in previous sections), which are gathered in Table \ref{tab:redbulkelastic}. We limit ourselves to 
the temperature interval where the $N_\text{TB}$ phase appears stable in the experiment (Table \ref{tab:temp}).
\begin{figure}[htp!]
    \centering
    \includegraphics[width=0.9\textwidth]{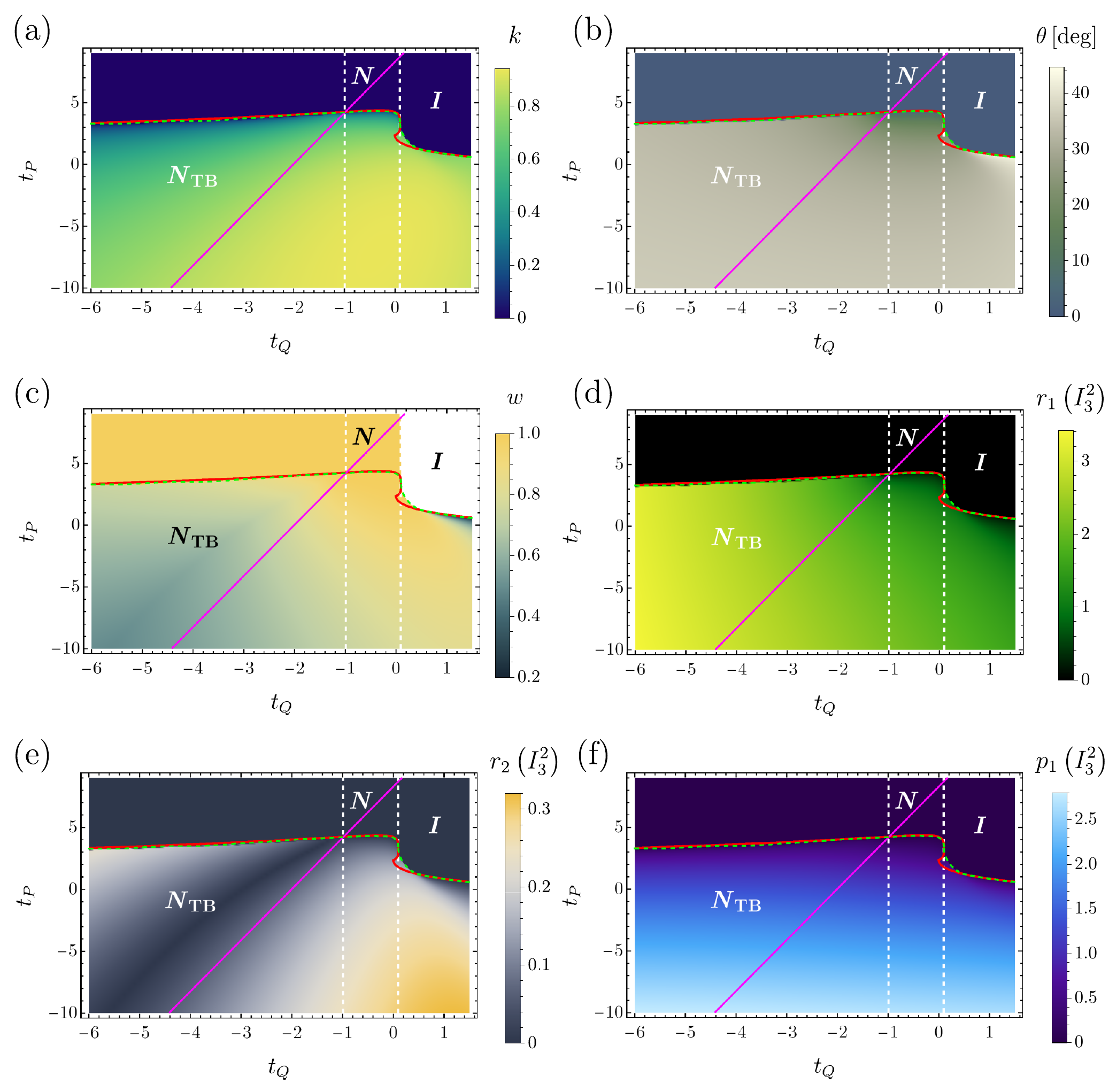}
    \caption{Phase diagram combined with density map of wave vector $k$ (a), tilt angle $\theta$ (b), biaxiality parameter $w$ (c) and mode's amplitudes: $r_1$ (d), $r_2$ (e) and $p_1$ (f) within theory $(I_3^2)$. Red continuous curve marks the bifurcation between $N\leftrightarrow N_\text{TB}$ and $I\leftrightarrow N_\text{TB}$, whereas dashed green curve outlines the numerical results. Magenta line (described by Eq.~(\ref{CB7CBexp})) reflects the phase sequence associated directly with the experimental data for CB7CB. Vertical, dashed white lines designate the temperature span of $N_\text{U}$ stability (experimental) mapped on $t_Q$ (\emph{see} Table \ref{tab:temp}).}
    \label{fig:densityplot}
\end{figure}
\begingroup
\renewcommand*{\arraystretch}{1.1}
\setlength{\tabcolsep}{12pt}
\begin{table}[h!]
\caption{Dimensionless parameters related to bulk part, elastic constants and bifurcation equations. Below are provided converters for $\Delta t \leftrightarrow t_Q$, $\tilde{\mathbf{r}}\leftrightarrow\mathbf{r}$ 
and $\tilde{\mathbf{k}}\leftrightarrow\mathbf{k}$.}
\begin{tabular}{cc||cc||cc}
\colrule
Parameter & Value&Parameter & Value& Parameter & Value\\
\colrule
$c_b$&0.67&$\rho_{2,2}$&$-0.0048$ &$l_2$&$-1.58$\\ 
$d_b$&-1.29&$\rho_{3,2}$& $-3.64$&$l_3$& $0.59$\\ 
$\Delta t_{NI}$&0.0029&$\rho_{3,3}$& $0.74$ &$l_4$& $0.93$\\ 
&&$\rho_{3,4}$& $0.29$ &$\kappa_1$& $1.13$\\ 
&&$\rho_{4,6}$&$0.62$ & $\kappa_2$& $5.38$\\
&&$\rho_{4,7}$& $0.15$ & $\kappa_3$& $-7.73$\\
&&$\rho_{4,14}$& $1.14$ & $\kappa_4$& $6.0054$\\ 
\colrule
\multicolumn{6}{c}{$t_Q=32.36(\Delta t+\Delta t_{NI})\quad\quad\tilde{\mathbf{r}}\,\text{[nm]}=1.06\, 
\mathbf{r}$\quad\quad$\tilde{\mathbf{k}}\,\text{[nm$^{-1}$]}=0.94\, \mathbf{k}$}
\end{tabular}
\label{tab:redbulkelastic}
\end{table}
\endgroup

\begingroup
\renewcommand*{\arraystretch}{1}
\begin{table}[htp!]
\caption{Temperature ranges in $\Delta t$ and $t_Q$ units for stable liquid crystalline structures: nematic and twist-bend nematic from experiment \cite{Babakhanova2017}.}
\begin{tabular}{cc}
\colrule
Range of temperature parameter & Description \\
\colrule
$-0.03<\Delta t\leq 0$  & nematic \\
$\Delta t\leq -0.03$  & twist-bend nematic\\
$-0.99 < t_Q \leq 0.09$  & nematic\\
$t_Q \leq -0.99$  & twist-bend nematic\\
\colrule
\end{tabular}
\label{tab:temp}
\end{table}
\endgroup
The temperatures $t_P$ an $t_Q$ are connected with absolute temperature $T$ of the 
system studied (\emph{see} Eqs.~(\ref{aQ},\ref{fP},\ref{scaledVariables})). 
Since $a_P>0$, $a_Q>0$, $T^\ast>T_P$ and $T>T_P$ any straight 
line in $\{t_Q,t_P\}$ plane with positive slope and negative $t_Q$–intercept 
represents a permissible physical system
with no polar order for $\tilde{\mathbf{Q}}=0$. Thus, we present the 
phase diagrams in the general $\{t_Q,t_P\}$ plane for a broader view. 
In our case, the experimental-related line has the form:
\begin{equation}
    t_P=4.13 t_Q+8.29.
    \label{CB7CBexp}
\end{equation}
Results of our in-depth analysis, profoundly reduced the number of adjustable parameters for CB7CB-like compounds 
to solely four $(\lambda, e_P, a_d,\,\text{and}\, e_b )$. From considerations 
related to the elastic constants, Table \ref{tab:fitbulkelastic}, it turns out that  $\Lambda_\text{QP}$  
(\emph{ipso facto} $\lambda$), responsible for globally polar structures,
is negligible, i.e. $\sqrt{|\lambda|}/e_P\approx 10^{-13}$. 
Thus, we set $\lambda=0$. Onward, we take $e_P=7.1$ and $a_d=0.75$ as the best values to reproduce the temperature dependance of $k$. 
For the bulk biaxial parameter $e_b$ we 
take two values: $e_b=0$ and $e_b=1/6$. 
If we recall Eq.~(\ref{LdG_bulk_Red}), there is a term $e_b\big(I_2^3-6I_3^2\big)+I_3^2$. 
If we set $e_b=0$, then it reduces to $I_3^2$, 
on the other hand when we set $e_b=1/6$ then we have only $I_2^3/6$. 
In the following discussion the first scenario $(e_b=0)$ will be referred 
to as theory $\big(I_3^2\big)$ and second one as theory $\big(I_2^3\big)$.
The  $\big(I_3^2\big)$ theory  will enhance phase biaxiality 
due to its tendency to lower equilibrium value of the $w$ parameter, Eq.~(\ref{wQz}), 
while the $\big(I_2^3\big)$ theory is promoting the  $w=\pm 1$ states 
through cubic and fifth-order terms in (\ref{LdG_bulk_Red}) \cite{AllenderLonga&2008}. 
In this latter case the biaxiality of $N_\text{TB}$ 
can only be induced by the elastic terms.

\begin{figure}[htp!]
    \centering
    \includegraphics[width=0.9\textwidth]{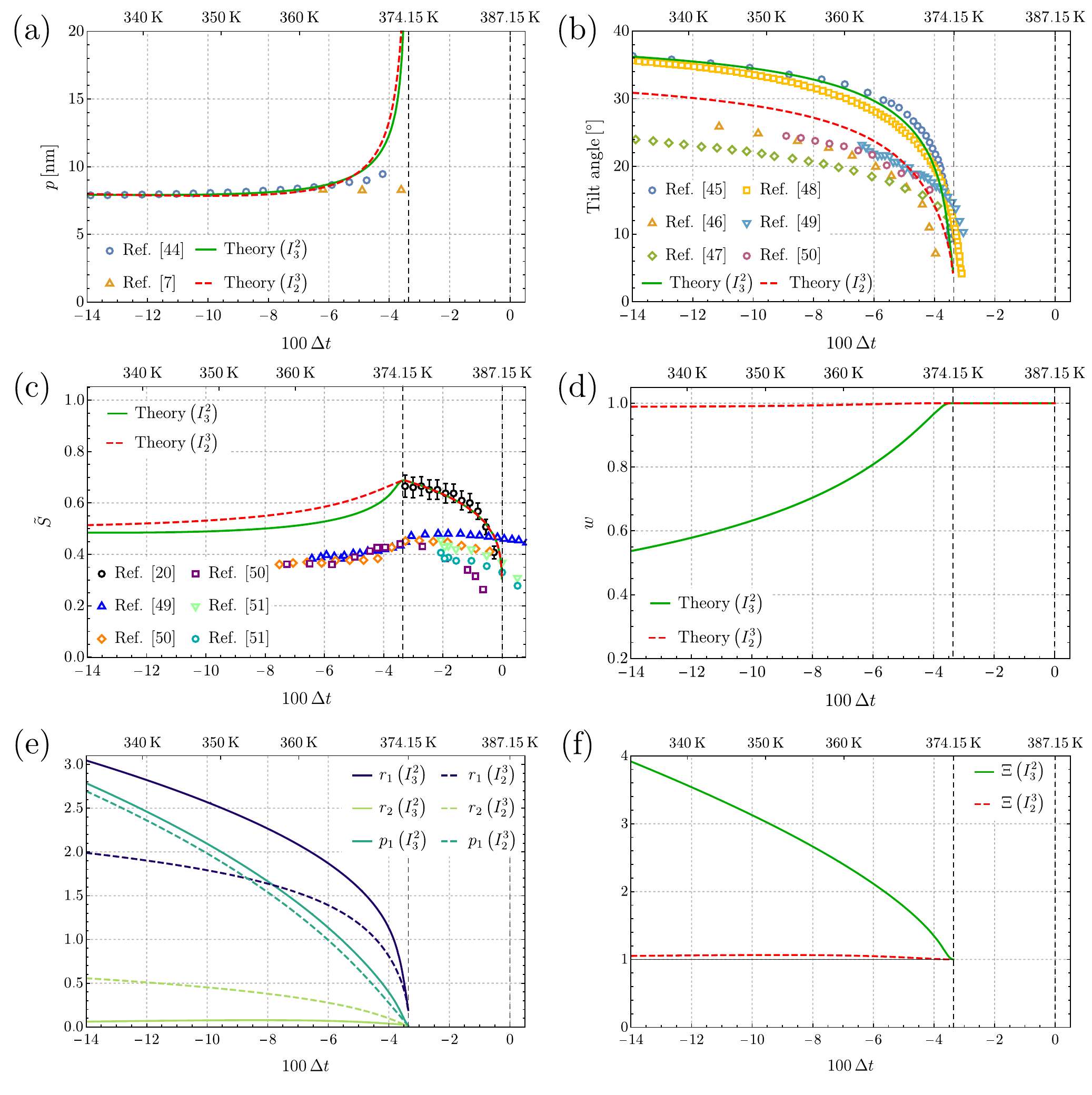}
    \caption{Comparison between experimental data (hollow points) and theoretical
    predictions (continuous green and dashed red line) for CB7CB's $N_\text{TB}$'s 
    temperature dependence of pitch $p$ (a), tilt angle $\theta$ (b) and order 
    parameter $\tilde{S}$ (c). Plot (d) depicts the behaviour of biaxiality 
    parameter $w$ and plot (e) mode's amplitudes: $r_1$, $r_2$ and $p_1$ as 
    a function of temperature in the range of $N_\text{TB}$ stability. Plot (f) 
    illustrates the temperature behaviour of the relevant factor parameterizing 
    the relative magnitudes of intensities of two leading harmonics of the dispersion tensor
    that contribute to the resonant soft X-ray scattering (RSoXS) \cite{supscat}. 
    All the data is drawn both with respect to the multiplied 
    by factor 100 reduced temperature $\Delta t$, 
    whereas key temperatures, corresponding to given $\Delta t$, 
    are designated above each plot in absolute temperature $T$.}
\label{fig:regularplot}
\end{figure}

Figs. \ref{fig:densityplot}a-f depict phase diagrams 
combined with density maps of $k,\theta$,  $w$, $r_1$, $r_2$ and $p_1$, respectively, which are outcomes of theory $(I_3^2)$.
In the analyzed case, being consistent with the experiment, 
stable, apart from isotropic, is the uniaxial nematic phase 
and the twist-bend nematic phase. Dashed green curve denotes 
numerically determined phase transitions and red continuous curve marks 
the results from the bifurcation analysis (\emph{see} Section V).  
Vertical, dashed white lines designate the temperature span 
of $N_\text{U}$ stability (experimental) mapped on $t_Q$ 
(\emph{see} Table \ref{tab:temp}). The purple straight line described 
by the Eq.~(\ref{CB7CBexp}) represents phase transition sequence:
$I\leftrightarrow N_\text{U}\leftrightarrow N_\text{TB}$
based on the CB7CB data from \cite{Babakhanova2017}. 
From points lying on this line we have attained information 
about the behavior of pitch $(p)$, tilt angle $(\theta)$ and 
nematic order parameter $(\tilde{S})$ in $N_\text{TB}$, alongside the 
insight into the $N_\text{TB}$'s 
biaxiality parameter $(w)$  and the remaining order parameters ($r_1$, $r_2$, $p_1$) 
(\emph{see} Figs.~\ref{fig:regularplot}a-e). With regard to the $w$ parameter, in the literature there are 
not available any results concerning the biaxiality of $N_\text{TB}$, 
thus it is hard to compare. Nevertheless, our model permits to estimate 
the span and magnitude of affect on experimentally measurable parameters. 

We set together results of our model with available experimental 
data concerning pitch $p$ (Fig.~6a \cite{ZhuChenhui2016,Chen2013}), tilt angle $\theta$ (Fig.~\ref{fig:regularplot}b  \cite{MeyerLuckhurstDozov2015,tuchb2017doublehelical,Jokisaari2015,Vaupotic2018,Mandle2019,Singh2016}) 
and nematic order parameter $\tilde{S}$ (Fig.~\ref{fig:regularplot}c  \cite{Babakhanova2017,Mandle2019,Singh2016,Zhang2015}). At the transition temperature from $N_\text{U}$ to $N_\text{TB}$ the pitch length is ca. 54 nm and with decreasing temperature it saturates at the level of ca. 8 nm (Fig. \ref{fig:regularplot}a). As one can see, it goes fairly well with the experimental data. Within the literature the methodology regarding the pitch measurements for CB7CB are consistent, i.e. all indicate that the pitch value reaches the plateau at ca. 8 nm \cite{Borshch2013,ZhuChenhui2016,Chen2013}, in contradiction to measurements of $\theta$ and $\tilde{S}$.

Such span of experimental data for $\theta$, Fig.~\ref{fig:regularplot}b, originate from the adopted method of determination and sample treatment. In Refs. \cite{MeyerLuckhurstDozov2015}, \cite{tuchb2017doublehelical} and \cite{Vaupotic2018} birefringence measurements were employed, however the choice of region in which they were taken varied across the aforementioned papers (\emph{see discussion in} \cite{tuchb2017doublehelical}). In Refs. \cite{Mandle2019} and \cite{Singh2016} the data regarding the conical tilt angle were extracted from X-ray methods, wherein in Ref. \cite{Mandle2019} it was small/wide angle X-ray scattering (SAXS/WAXS) and in Ref. \cite{Singh2016} X-ray diffraction (XRD). In turn, conical tilt angle from Ref. \cite{Jokisaari2015} was determined \emph{i.a.} from $^2$H Nuclear Magnetic Resonance (NMR) quadrupolar splittings of CB7CB-d$_4$.
Similarly to tilt angle, discrepancies between the data related to order parameter, Fig.~\ref{fig:regularplot}c, arise from the method of acquisition. In Ref. \cite{Babakhanova2017} 
it was extracted from diamagnetic anisotropy measurements, 
in Ref. \cite{Mandle2019} from SAXS, 
in Ref. \cite{Singh2016} from XRD and 
in Refs. \cite{Singh2016,Zhang2015} from polarized Raman spectroscopy (PRS). As one can see data from Ref. \cite{Babakhanova2017} stands out from the rest of the data (Fig.~\ref{fig:regularplot}c), although it was the only source that provided simultaneously the data for the temperature dependence of orientational order parameter and elastic constants.   \\
One can see that results of our model are generally in a very good agreement 
with experimental results, perhaps except an immediate vicinity of the 
$N_\text{U}-N_\text{TB}$ phase transition where fluctuations - not included in the present
analysis - may play a role. Interesting seem predictions concerning the 
effect of intrinsic, molecular biaxiality on $N_\text{TB}$. While pitch, $\tilde{S}$
and $p_1$ are practically insensitive to $w$ the remaining observables 
are affected. In particular, for tilt angle the green continuous line, 
associated with $e_b=0$ fits well in between the data from Refs. \cite{MeyerLuckhurstDozov2015} 
(blue circles) and \cite{Vaupotic2018} (yellow squares), 
whereas the red dashed line, associated with the weakly biaxial 
case, markedly departs from the above experimental data.
Based on the results of Babakhankova \emph{et al.} \cite{Babakhanova2017}
we can conclude  that biaxiality of $N_\text{TB}$, initially small at the 
$N_\text{U}-N_\text{TB}$ phase transition, considerably increases on departing 
from the transition temperature  (green line in Fig.~\ref{fig:regularplot}d).   
Fig. \ref{fig:regularplot}e illustrates the behavior of the order parameters $r_1$, 
$r_2$ and $p_1$ in $N_\text{TB}$ phase of CB7CB, where the ratio  $\sigma=r_1/r_2$ 
can be correlated with data acquirable from resonant soft X-ray scattering (RSoXS)  
\cite{ZhuChenhui2016,Salamonczyk2017,Salamonczyk2018}.
In order to make this correlation, we translated our formalism 
into the one presented in \cite{supscat}. Thanks to that, we could tie the
results for $\sigma$ with experimentally measurable scattering intensities 
through following formula:
\begin{equation}
    \Xi=-\frac{1}{2}+\frac{3\sigma}{2\sigma\cos(2\theta)+4\sin(2\theta)},
\end{equation}
where $\Xi=f_1/f_2$ and  $\theta$ is a conical tilt angle. 
The value of parameter $\Xi$ determines the intensity of the $2q_0$ 
peak (half-pitch band) with the respect to the intensity of the 
$q_0$ peak (full pitch band) in the $N_\text{TB}$ phase, where $q_0=2\pi/p$ 
is is the magnitude of the wave vector of the heliconical deformation 
with the pitch $p$. As it was stated in \cite{supscat}, if $\Xi\geq 1$ 
then the intensity of the $2q_0$ peak is approximately two orders of 
magnitude lower than the intensity of the $q_0$ peak and further strongly 
decreases with increasing $\Xi$. On the other hand, if $\Xi< 1$ then the 
intensity of the $2q_0$ peak escalates rapidly. As one can see in 
Fig.~\ref{fig:regularplot}f, which illustrates the temperature 
dependence of $\Xi$ for both theories $I_3^2$ and $I_2^3$, 
all the data obeys the relation $\Xi\geq 1$ indicating a 
significant weakness of $2q_0$ peak. Interestingly,
while for the  $I_2^3$ theory the relative magnitude of the 
intensities should roughly differ by two orders of magnitude irrespective 
of the temperature the  $I_3^2$ model predicts further strong reduction
of the relative intensity with temperature. To our best knowledge,
the $2q_0$ signal has not been detected so far in any of the examined 
$N_\text{TB}$ forming compounds \cite{ZhuChenhui2016,Stevenson2017,Salamonczyk2017,Salamonczyk2018}.
\section{Conclusions}
Currently existing experimental data 
are in favour of the theory that the $N_\text{U}-N_\text{TB}$ 
phase transition is driven by the flexopolarization mechanism.
According to this theory deformations of the director induce
local  polar order which, in turn, 
renormalizes the bend elastic constant to a very small value 
relative to other elastic constants, eventually leading to the 
twist-bend instability. However, a fundamental description of 
orientational properties of nematics based on minimal coupling 
Landau-de Gennes  theory of flexopolarization
suggests that  such  behavior does not need to be universal.  Even 
when both $K_{11}$ and $K_{33}$ are simultaneously reduced due to 
splay-bend degeneracy (inherent to the minimal coupling LdeG expansion)
the $N_\text{TB}$ phase can still become absolutely stable 
among all possible one dimensional periodic structures.
Since this case has not been observed experimentally to date
an important question that arises is whether  
the flexopolarization mechanism is indeed sufficient to 
explain the experimental observations at the level of 
the ``first principles'' Landau-deGennes theory of orientational 
order. In this work we gave a positive answer to this question. 
We demonstrated that the experimental observations 
involving the nematic twist-bend phase 
and the related  uniaxial nematic phase 
can be explained if we generalize the minimal coupling
theory to the level where the properties of high-temperature
uniaxial nematic phase are properly accounted for.
Especially, the constructed generalized free energy density 
is in line with experimentally observed  temperature variation 
of the Frank elastic constants and of the orientational order 
parameter.  

Our proposed generalized theory of uniformly distorted nematics extends 
the elastic part of LdeG  by additional two terms of third-order.
The added elastic terms are the only independent ones
for UDF and depending on model parameters various UDF 
structures can become minimizers
of the free energy, including nematic twist-bend. This conclusion 
follows directly from the bifurcation analysis and from the observation
that the remaining four independent elastic  
terms, not included in the theory, 
can always be written in such a way that they vanish for UDF. 
It is worth noticing that only one more term
is generally needed to extend the studies of the UDF class to 
\emph{all possible} one-dimensional distortions of the alignment tensor. 
This sort of classification of various elastic terms is similar to 
what Virga has recently proposed  
for generalized Frank’s elastic theory \cite{Virga2019}.

The numerical analysis of the model quite well reproduces
measured quantities for the $N_\text{U}$ and  $N_\text{TB}$ phases
of the CB7CB-like mesogens and gives numerical estimates 
for its constitutive parameters including 
otherwise difficult to access (flexo-)polarization couplings.
Overall, the $N_\text{TB}$ phase is predicted to be biaxial with theoretical 
support that major contribution to the phase biaxiality 
comes from bulk terms in the free energy.
The phase transition to the $N_\text{TB}$ is weakly first-order
although, in general, the theory permits the transitions 
to $N_\text{TB}$ be second order with tricritical $I-N_\text{TB}$ 
and $N_\text{U}-N_\text{TB}$ points.
 
Summarizing, the theory successfully links 
the  chiral order of $N_\text{TB}$ with the
flexopolarization mechanism. 
It can also serve as a starting point in seeking for a new 
classes of modulated nematic structures, like nematic splay or 
nematic splay-bend ones.
\begin{acknowledgments}
This work was partly supported by the Grant No. DEC-2013/11/B/ST3/04247 
of the National Science Centre in Poland. {LL acknowledges partial support 
by the EPSRC grant EP/R014604/1 [The Mathematical Design of New Materials-2019] 
of the Isaac Newton Institute for Mathematical Sciences, University of Cambridge, UK.} 
WT acknowledges a partial support by Marian Smoluchowski Scholarship 
(KNOW/58/SS/WT/2016) from Marian Smoluchowski Cracow Scientific 
Consortium ``Matter-Energy-Future'' within the KNOW grant and by 
Jagiellonian Interdisciplinary PhD Programme co-financed from the European 
Union funds under the European Social Fund (POWR.03.05.00-00-Z309/17-00). 
{Results from this work were  presented during the 2019 Gordon Conference 
on Liquid Crystals (July 7 - 12, New London, NH, USA) and 
the Erwin Schr\"{o}dinger Institute (ESI) Workshop: New trends in 
the variational modeling and simulation of liquid crystals 
(December 2 - 6, Vienna, Austria, 2019).}
\end{acknowledgments}
\bibliographystyle{apsrev4-1}
\bibliography{References}
\end{document}